\documentclass[preprint,amsmath,amssymb,aps]{revtex4-1}

\usepackage{graphicx}
\usepackage{dcolumn}
\usepackage{bm}
\usepackage{epsfig}
\usepackage{epstopdf}
\usepackage{subfig}
\usepackage{caption}
\usepackage{enumerate}
\usepackage{comment}
\usepackage{physics}
\usepackage{amsmath}
\usepackage{slashed}
\usepackage{ulem}
\usepackage{color}
\usepackage{hyperref}
\usepackage [autostyle, english = american]{csquotes}
\MakeOuterQuote{"}

\graphicspath{{images/}}
\newcommand{\Lagr}{\mathcal{L}}

\begin{document}

\title{Infrared divergences in $e^{+}e^{-}\rightarrow 2\; \text{jets}$ in the light front coherent state formalism}

\author{Deepesh Bhamre}
 \email{deepesh.bhamre@physics.mu.ac.in}
 \affiliation{Department of Physics, University of Mumbai, Santacruz (East), Mumbai-400098, India}
\author{Anuradha Misra}
 \email{Formerly Department of Physics, University of Mumbai, Santacruz (East), Mumbai-400098, India misra@physics.mu.ac.in}
 \affiliation{Centre for Excellence in Theoretical and Computational Sciences (CETACS), University of Mumbai, Santacruz (East), Mumbai-400098, India}

\begin{abstract}
	We study infrared (IR) divergences in light front quantum chromodynamics using a coherent state basis in light front time-ordered Hamiltonian perturbation theory. In computation of the S-matrix elements in Hamiltonian formalism, the IR divergences  appear in the form of vanishing energy denominators. We consider the process $e^{+}e^{-}\rightarrow 2\; \text{jets}$ at $\mathcal{O}(g^{2})$ in strong coupling, construct the coherent state representing the outgoing particles, and explicitly show that the `true' IR divergences cancel to this order when the matrix elements are calculated between coherent states instead of Fock states.

\begin{description}

\item[Keywords]
{Infrared divergences, Coherent State Formalism, Light Front QCD}

\end{description}

\end{abstract}

\maketitle

\section{Introduction}\label{sec:intro}

Infrared (IR) divergences in quantum electrodynamics (QED) and quantum chromodynamics (QCD) are known to cancel for infrared safe observables when contributions of real emissions are added to virtual corrections, which essentially amounts to summing over physically indistinguishable states in the cross section calculations. This real+virtual cancellation takes place at the level of cross section and is well established as a successful approach to the IR problem. 

An alternative approach, proposed originally for QED by Kulish and Faddeev (referred to as KF approach henceforth), is based on the method of asymptotic dynamics \cite{kulishfaddeev} and  addresses the issue of cancellation of IR divergences at the amplitude level. The KF approach is based on Hamiltonian perturbation theory wherein the S-matrix elements are obtained by calculating the matrix elements between coherent states, which are the asymptotic states obtained by applying the asymptotic evolution operator to the Fock states. The rationale behind the use of coherent states in place of Fock states, in the calculation of S-matrix elements, is that for theories containing long-range interactions (as also for theories like QCD where the initial and final states are bound states), the underlying assumption in the Lehmann, Symanzik and Zimmermann formalism - that the asymptotic states are eigenstates of free particle Hamiltonian - is no longer valid. The method of asymptotic dynamics is based on the observation that the interaction Hamiltonian does not approach zero at the asymptotic limits ($t \rightarrow \pm \infty$), and hence the correct asymptotic states need to be obtained by applying the "asymptotic evolution operator" to the Fock states. The asymptotic evolution operator contains the asymptotic Hamiltonian which is obtained by taking the limiting value of interaction Hamiltonian at infinite times. It was shown by Kulish and Faddeev that the IR divergences  in QED cancel at the amplitude level itself when the S-matrix elements are calculated between coherent states constructed in this manner, instead of the Fock states. The coherent state method was subsequently discussed in the context of QCD also (see Refs.  \cite{catanigeneralizedcohst,cataninoncancellingir,catanigaugecovariance,ciafalonicohstasydyn,ciafaloniirsingcohst,delduca,fordesigner,hannesdottirfinitesmatrix,hannesdottirmassless}).

A coherent state approach, in the context of light front QED (LFQED), was developed by one of us in Ref.~\cite{misraqed} and was shown to lead to cancellation of IR divergences in the one-loop vertex correction in QED. Application of the coherent state basis to the discretized light cone quantization (DLCQ) method of bound state calculations was discussed in \cite{misradlcq} and preliminary study of coherent states in light front QCD (LFQCD) was performed in Refs.~\cite{misraqcd} and \cite{misrafbs}. IR divergences in fermion self-energy correction in LFQED were discussed in Refs.~\cite{jaimisra,jaimisrafeynmangauge,jaimisraallorder} and it was shown that the IR divergences in fermion self-energy correction cancel to all orders at amplitude level in LFQED. 

In this work, we focus on the construction of coherent states and cancellation of IR divergences in coherent state basis in LFQCD. In this preliminary study, we  consider the process $e^{+}e^{-}\rightarrow 2\; \text{jets}$ to $\mathcal{O}(g^{2})$ in strong coupling and show by explicit calculations the cancellation of `true' IR divergences. In the coherent state approach, a transition amplitude receives, in addition to the conventional Fock state contributions that are IR-divergent, contributions from higher-order terms of the coherent state. These too are IR-divergent by construction since the phase space is limited to the region where only those particles that are either soft or are collinear to external particles are emitted/absorbed. These two IR-divergent contributions are expected to cancel between themselves leading to an IR-finite transition amplitude. A somewhat similar study has been performed by Forde and Signer in equal-time QCD in Ref.~\cite{fordesigner}, wherein the authors have also discussed the advantages and practical use of an amplitude approach as compared to the cross section approach.

A brief introduction to coherent states in massless QFTs and of recent developments is provided in Ref.~\cite{magneairreview}. Unless an explicit distinction between the `soft' and `collinear' divergences is made, the generic term `infrared divergences' will be used to denote both in this paper.

In Sec.\ref{sec:lfqcd}, we review the derivation of LFQCD Hamiltonian presented in Ref.~\cite{brodskyreview}. In Sec.\ref{sec:coherent_states}, a brief account of IR divergences appearing in LFQCD is given and the coherent state formalism is described. Further, the coherent state, to the order necessary for our calculation, is constructed in this section. The asymptotic region of phase space which leads to IR divergences and its relevance in the construction of coherent states is also described. In  Sec.\ref{sec:cancellation}, we show the cancellation of IR divergences at $\mathcal{O}(g^{2})$ in the coherent state formalism for the process $e^{+}e^{-}\rightarrow 2\; \text{jets}$, and conclude in Sec.\ref{sec:summary}. Appendix \ref{app:Amplitude_calculation} contains a detailed calculation of one of the terms of the amplitude, primarily to illustrate the technique of light cone time-ordered perturbation theory being used in our work.

\section{Light Front QCD Hamiltonian}\label{sec:lfqcd}
In this section, we review the derivation of LFQCD Hamiltonian following  Ref.~\cite{brodskyreview}. The standard QCD Lagrangian (without the gauge-fixing terms) is given by
\begin{equation} \label{eq:QCD_Lagrangian}
\Lagr=-\frac{1}{4}F^{\mu\nu}_{a}F_{\mu\nu}^{a}+\frac{1}{2}[\bar{\Psi}_{c}(i\gamma^{\mu}D_{\mu}^{cc'}-m\delta_{cc'})\Psi_{c'}+H.c.],
\end{equation}
where $F^{\mu\nu}_{a}$ is the field tensor given by
\begin{equation} \label{eq:field_tensor}
F^{\mu\nu}_{a}=\partial^{\mu}A^{\nu}_{a}-\partial^{\nu}A^{\mu}_{a}-gf^{abd}A^{\mu}_{b}A^{\nu}_{d}.
\end{equation}
The fermionic (quark) fields in Eq.(\ref{eq:QCD_Lagrangian}) carry color indices $c,c'=1,2,3$. $A^{\mu}_{a}$ in Eq.(\ref{eq:field_tensor}) are eight real-valued color vector potentials with $a=1,2,...,8$ being the gluon index. The Hermitian, traceless 3 x 3 matrix
\begin{equation}
A^{\mu}=\frac{1}{2}
\begin{bmatrix}
\frac{1}{\sqrt{3}}A^{\mu}_{8}+A^{\mu}_{3} & A^{\mu}_{1}-iA^{\mu}_{2} & A^{\mu}_{4}-iA^{\mu}_{5} \\
A^{\mu}_{1}+iA^{\mu}_{2} & \frac{1}{\sqrt{3}}A^{\mu}_{8}-A^{\mu}_{3} & A^{\mu}_{6}-iA^{\mu}_{7} \\
A^{\mu}_{4}+iA^{\mu}_{5} & A^{\mu}_{6}+iA^{\mu}_{7} & -\frac{2}{\sqrt{3}}A^{\mu}_{8} \\
\end{bmatrix}\\
\end{equation}
belongs to the $SU(3)_{color}$ gauge group. $A^{\mu}$ is constructed using the definition
\begin{equation}
(A^{\mu}(x))_{cc'}\equiv T^{a}_{cc'}A^{\mu}_{a}(x),
\end{equation}
where the color matrices $T^a$ are related to Gell-Mann matrices $\lambda^{a}$ through
\begin{equation}
T^{a}=\frac{1}{2}\lambda^{a}.
\end{equation}
The color matrices obey the relations
\begin{equation}
[T^{a},T^{b}]_{cc'}=if^{abd}T^{d}_{cc'};\;\;Tr(T^{a}T^{b})=\frac{1}{2}\delta^{ab}.
\end{equation}
The covariant derivative appearing in Eq.(\ref{eq:QCD_Lagrangian}) is given by
\begin{equation}
D^{\mu}_{cc'}=\partial^{\mu}\delta_{cc'}+igT^{a}_{cc'}A^{\mu}_{a}.
\end{equation}
The Hamiltonian can be obtained from the energy-momentum stress tensor $T^{\mu\nu}$. The light cone Hamiltonian is
\begin{equation} \label{eq:raw_QCD_Hamiltonian}
\begin{split}
P^{-}=P_{+}=& \int d^{2}{\bf{x}}_{\perp}dx^{-} T^{+-}\\
=& \int d^{2}{\bf{x}}_{\bf\perp}dx^{-}\bigg(F^{+\kappa}_{a}F_{\kappa +}^{a}+\frac{1}{4}F^{\kappa\lambda}_{a}F_{\kappa\lambda}^{a}+\frac{1}{2}[i\bar{\Psi}\gamma^{+}T^{a}D^{a}_{+}\Psi + H.c.]\bigg)
\end{split}
\end{equation}

\begin{figure}[ht]
\centering
\subfloat[$V_{1}$]
{\includegraphics[width=0.25\columnwidth]{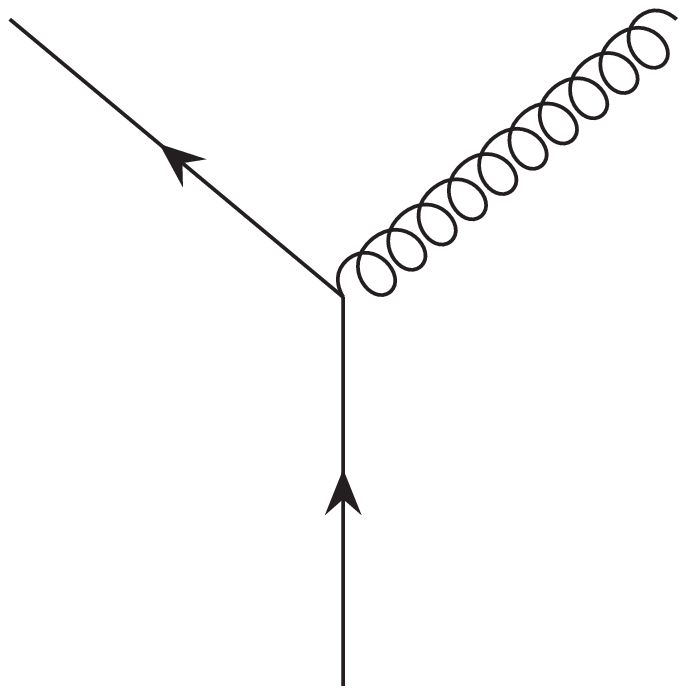}}
\hspace{1cm}
\subfloat[$V_{2}$]
{\includegraphics[width=0.25\columnwidth]{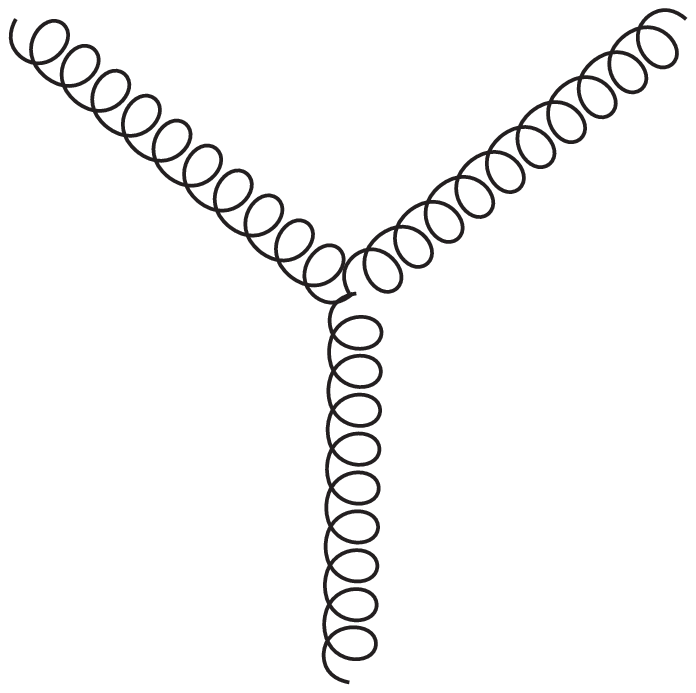}}
\hspace{1cm}
\subfloat[$V_{3}$]
{\includegraphics[width=0.3\columnwidth]{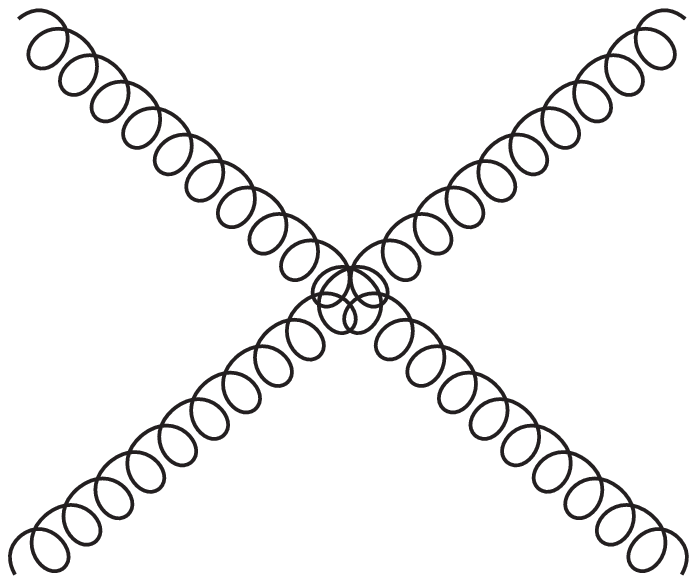}}
\hspace{1cm}
\subfloat[$W_{1}$]
{\includegraphics[width=0.4\columnwidth]{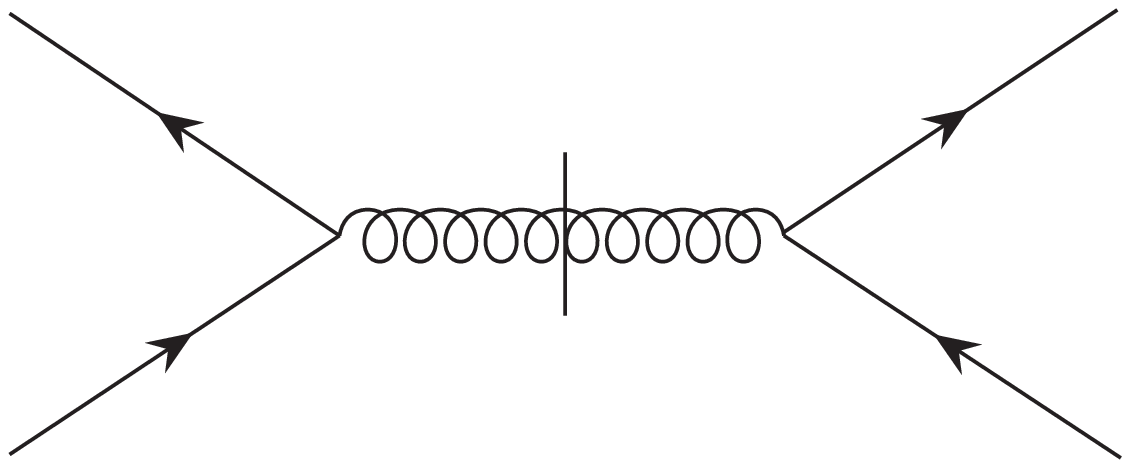}}
\hspace{1.5cm}
\subfloat[$W_{2}$]
{\includegraphics[width=0.4\columnwidth]{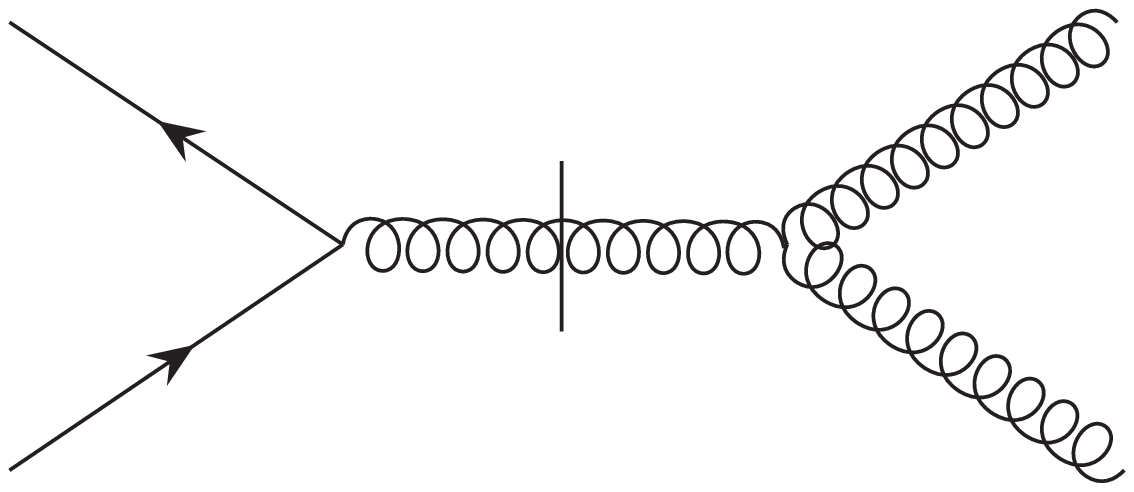}}
\hspace{1cm}
\subfloat[$W_{3}$]
{\includegraphics[width=0.4\columnwidth]{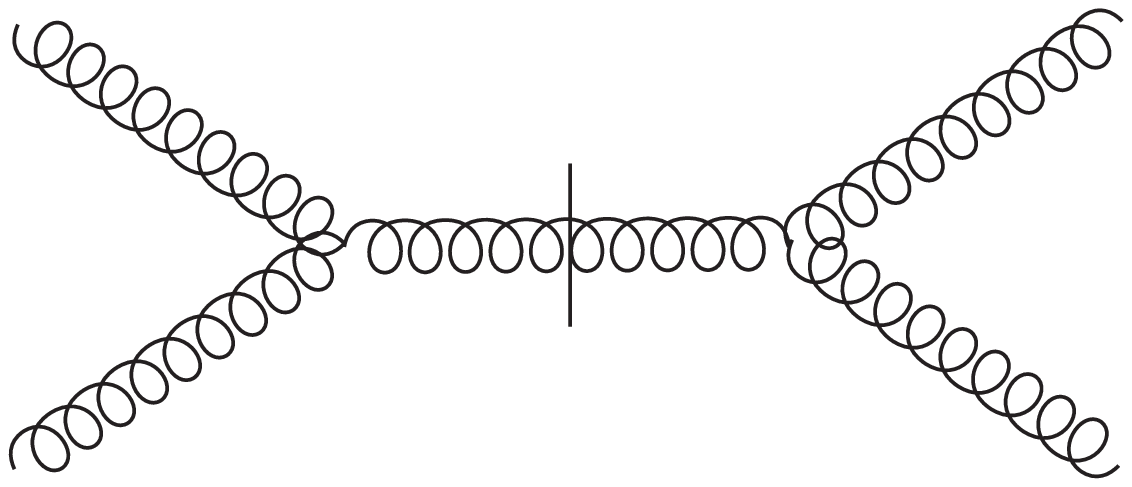}}
\hspace{1.5cm}
\subfloat[$W_{4}$]
{\includegraphics[width=0.4\columnwidth]{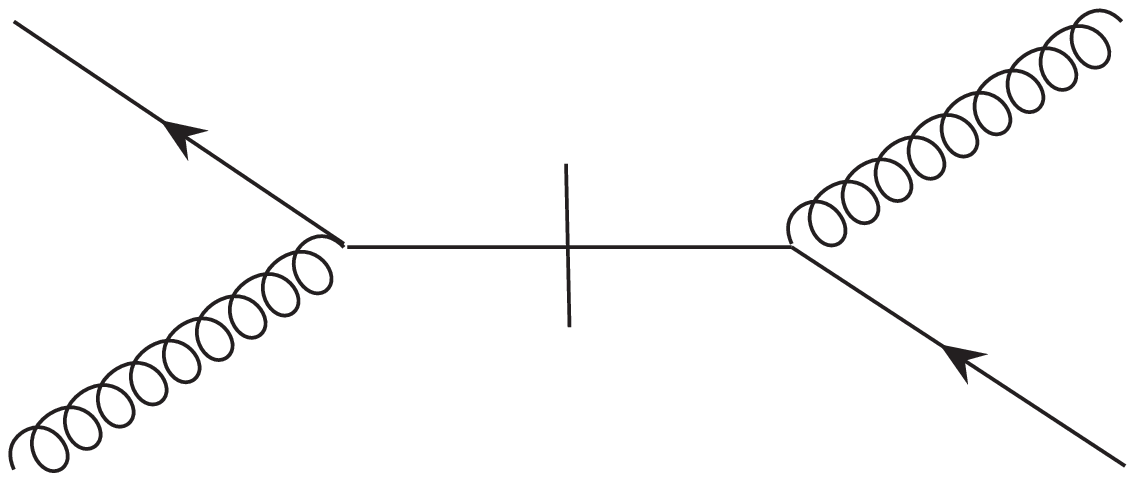}}
\caption{Vertices in LFQCD}
\label{fig:vertices_QCD}
\end{figure}

Using light cone gauge $A^{+}_{a}=0$, it is seen that the Euler-Lagrange equations
\begin{equation} \label{eq:constraint_gluon}
-\partial^{+}\partial_{-}A^{-}_{a}-\partial^{+}\partial_{i}A^{i}_{\perp a}=gJ^{+}_{a}
\end{equation}
and
\begin{equation} \label{eq:constraint_fermion}
2i\partial_{-}\Psi_{-}=(m\beta-i\alpha^{i}_{\perp}T^{a}D^{a}_{\perp i})\Psi_{+}
\end{equation}
are constraint equations. Here $\alpha^{k}=\gamma^{0}\gamma^{k}$, $\beta=\gamma^{0}$, and $\Psi_{\pm}=\Lambda_{\pm}\Psi$ are spinor projections obtained using the operators $\Lambda_{\pm}=\frac{1}{2}(1\pm\alpha^{3})$. The dependent degrees of freedom $A^{-}_{a}$ and $\Psi_{-}$ are eliminated by inverting Eqs.(\ref{eq:constraint_gluon}) and (\ref{eq:constraint_fermion}) respectively. The free solutions, in terms of the independent degrees of freedom, are  
\begin{equation}
\tilde{A}^{\mu}_{a}=\bigg(0,-\frac{1}{\partial^{+}}\partial_{i}A^{i}_{a},A_{\perp}^{a}\bigg)
\end{equation}
and
\begin{equation}
\begin{split}
\tilde{\Psi}=&\tilde{\Psi}_{+}+\tilde{\Psi}_{-}\\
=&\tilde{\Psi}_{+}+\frac{(m\beta-i\alpha^{i}\partial_{i})}{2i\partial_{-}}\tilde{\Psi}_{+}.
\end{split}
\end{equation}
Thus, the Hamiltonian in Eq.(\ref{eq:raw_QCD_Hamiltonian}), now rewritten in terms of the free solutions and thereby only the independent degrees of freedom in the theory, is given by
\begin{equation}\label{eq:LFQCD_Hamiltonian}
\begin{split}
P^{-}= \int d^{2}{\bf{x}}_{\perp}dx^{-}\bigg[ &\frac{1}{2}\bar{\Psi}\gamma^{+}\frac{m^{2}+(i\nabla_{\perp})^{2}}{(i\partial^{+})}\Psi+\frac{1}{2}A^{\mu}_{a}(i\nabla_{\perp})^{2}A_{\mu}^{a}\\
&+gJ^{\mu}_{a}A_{\mu}^{a}+\frac{g^{2}}{4}B^{\mu\nu}_{a}B_{\mu\nu}^{a}+\frac{g^{2}}{2}J^{+}_{a}\frac{1}{(i\partial^{+})^{2}}J^{+}_{a}\\
&+\frac{g^{2}}{2}(\bar{\Psi}\gamma^{\mu}T^{a}A_{\mu}^{a})\frac{\gamma^{+}}{(i\partial^{+})}(\gamma^{\nu}T^{b}A_{\nu}^{b}\Psi)\bigg],
\end{split}
\end{equation}
where we have discarded the tilde notation i.e. $\tilde{A}^{\mu}_{a}\rightarrow A^{\mu}_{a}$ and $\tilde{\Psi}_{+}\rightarrow \Psi_{+}$ from Eq.(\ref{eq:LFQCD_Hamiltonian}) onwards. In Eq.(\ref{eq:LFQCD_Hamiltonian}), $J^{\mu}_{a}=\bar{\Psi}_{c}\gamma^{\mu}T^{a}_{cc'}\Psi_{c'}+f^{abd}F^{\mu\alpha}_{b}A_{\alpha}^{d}$ is the color Maxwell current and $B^{\mu\nu}_{a}\equiv f^{abc}A^{\mu}_{b}A^{\nu}_{c}$. The first two terms in Eq.(\ref{eq:LFQCD_Hamiltonian}) form the free part of the Hamiltonian. Rest of the terms contain the strong coupling constant `$g$' and are the interaction terms of the theory. Using the expressions for $J^{\mu}_{a}$ and $B^{\mu\nu}_{a}$, the interaction part of the Hamiltonian is expanded to obtain the interaction Hamiltonian,
\begin{equation} \label{eq:QCD_Hamiltonian_splitup}
H_{\text{int}}=V_{1}+V_{2}+V_{3}+W_{1}+W_{2}+W_{3}+W_{4},
\end{equation}
where
\begin{align} \label{eq:V_1}
V_{1}&\equiv \int d^{2}{\bf{x}}_{\perp}dx^{-} g\bar{\Psi}\gamma^{\mu}T^{a}\Psi A^{a}_{\mu}
\\
V_{2}&\equiv \int d^{2}{\bf{x}}_{\perp}dx^{-} gf^{abd}(\partial^{\mu}A^{\nu}_{b})A_{\nu}^{d}A_{\mu}^{a}
\\
V_{3}&\equiv \int d^{2}{\bf{x}}_{\perp}dx^{-}\frac{g^{2}}{4}f^{abd}f^{aef}A^{\mu}_{b}A^{\nu}_{d}A_{\mu}^{e}A_{\nu}^{f}
\\ \label{eq:W_1}
W_{1}&\equiv \int d^{2}{\bf{x}}_{\perp}dx^{-} \frac{g^{2}}{2}(\bar{\Psi}\gamma^{+}T^{a}\Psi)\frac{1}{(i\partial^{+})^{2}}(\bar{\Psi}\gamma^{+}T^{a}\Psi)
\\ \nonumber
W_{2}&\equiv \int d^{2}{\bf{x}}_{\perp}dx^{-} \frac{g^{2}}{2}\bigg[(\bar{\Psi}\gamma^{+}T^{a}\Psi)\frac{1}{(i\partial^{+})^{2}}f^{abd}(\partial^{+}A^{\mu}_{b})A_{\mu}^{d}\\
& ~~~~~~~~~~~~~~~~~~~~~~~~~+f^{abd}(\partial^{+}A^{\mu}_{b})A_{\mu}^{d}\frac{1}{(i\partial^{+})^{2}}(\bar{\Psi}\gamma^{+}T^{a}\Psi)\bigg]
\\
W_{3}&\equiv \int d^{2}{\bf{x}}_{\perp}dx^{-} \frac{g^{2}}{2}f^{abd}f^{aef}(\partial^{+}A^{\mu}_{b})A_{\mu}^{d}\frac{1}{(i\partial^{+})^{2}}(\partial^{+}A^{\nu}_{e})A_{\nu}^{f}
\\
W_{4}&\equiv \int d^{2}{\bf{x}}_{\perp}dx^{-} \frac{g^{2}}{2}(\bar{\Psi}\gamma^{\mu}T^{a}A_{\mu}^{a})\frac{\gamma^{+}}{(i\partial^{+})}(\gamma^{\nu}T^{b}A_{\nu}^{b}\Psi).
\end{align}
In addition to the usual interactions in equal-time QCD, which are denoted by $V$ in Eq.(\ref{eq:QCD_Hamiltonian_splitup}), the interaction Hamiltonian also contains nonlocal interactions denoted here by $W$. The presence of the $W$ type i.e., instantaneous interaction terms is a result of eliminating the dependent degrees of freedom from the theory. Each of the interaction terms above is represented by a vertex as shown in Fig.\ref{fig:vertices_QCD}. Propagators with small vertical lines in Figs.\ref{fig:vertices_QCD}(d) through \ref{fig:vertices_QCD}(g) denote nonlocal (instantaneous) interactions.

The momentum-space expansion of the fields appearing in the Hamiltonian is
\begin{equation} \label{eq:Fourier_fields}
\begin{split}
&\Psi_{cf}(x)=\sum_{\lambda}\int[dp](b_{p\lambda cf}u_{p\lambda cf}e^{-ip{\cdot}x}+d^{\dagger}_{p\lambda cf}v_{p\lambda cf}e^{ip{\cdot}x})\\
&A^{a}_{\mu}(x)=\sum_{\lambda}\int[dp](a_{p\lambda a}\epsilon_{p\lambda a}e^{-ip{\cdot}x}+a^{\dagger}_{p\lambda a}\epsilon^{*}_{p\lambda a}e^{ip{\cdot}x}),
\end{split}
\end{equation}
where $\lambda$ is the spin (polarization) of the fermion (boson) and the color index is given by $c\ (a)$ for the fermion (boson). The fermion also carries the index $f$ for flavor quantum number. The light front integral measure is
\begin{equation}
\int[dp]\equiv \int_{-\infty}^{\infty}\frac{d^{2}{\bf{p}}_{\perp}}{(2\pi)^{3/2}}\int_{0}^{\infty}\frac{dp^{+}}{\sqrt{2p^{+}}}
\end{equation}
and the commutation (anti-commutation) relations for bosons (fermions) are
\begin{equation}
[a_{p\lambda a},a^{\dagger}_{p'{\lambda}' a'}]=\delta(p^{+}-p'^{+})\delta^{2}({\bf{p}}_{\perp}-{\bf{p'}}_{\perp})\delta_{\lambda}^{\lambda^{'}}\delta_{a}^{a'},
\end{equation}
\begin{equation}
\{b_{p\lambda cf},b^{\dagger}_{p'{\lambda}' c'f'}\}=\{d_{p\lambda cf},d^{\dagger}_{p'{\lambda}' c'f'}\}=\delta(p^{+}-p'^{+})\delta^{2}({\bf{p}}_{\perp}-{\bf{p'}}_{\perp})\delta_{\lambda}^{\lambda^{'}}\delta_{c}^{c'}\delta_{f}^{f'},
\end{equation}
respectively, with the rest of the commutators (anticommutators) being zero.

\section{IR divergences in LFQCD and the coherent state formalism}\label{sec:coherent_states}

\begin{figure}[ht]
\centering
{\includegraphics[width=0.5\columnwidth]{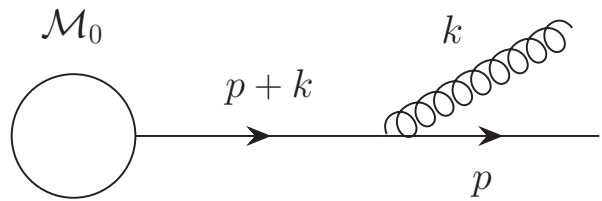}}
\caption{IR divergence due to emission of a soft or collinear gluon}
\label{fig:IR_div_gluon}
\end{figure}

It is well known that IR divergences occur when massless propagators go on shell. In addition to the `soft' divergences, gauge theories with a vertex involving all massless particles contain another type of divergence called `collinear' divergence. For example, consider a process in which a final-state massless quark is accompanied by a gluon, as shown in Fig.\ref{fig:IR_div_gluon}. The quark propagator has momentum $p+k$ and can become on shell in the following situations:\\
(i) $k^{\mu}\rightarrow 0\implies (p+k)^{2}\rightarrow p^{2}=0$ leading to soft divergence.\\
(ii) $k^{\mu}\parallel (p+k)^{\mu}\implies k^{\mu}=\lambda(p+k)^{\mu} \implies (p+k)^{2}\rightarrow (1+\lambda)^{2}p^{2}=0$ leading to collinear divergence.

In the light front formalism, the condition ${\bf{k}}=(k^{+},{\bf{k}}_{\perp})\rightarrow \bf{0}$ could either lead to a `spurious' or a `true' IR divergence. The dispersion relation in light front coordinates is given by $k^{-}=\frac{{\bf{k}}_{\perp}^{2}}{2k^{+}}$ for massless particles. Due to this form of the mass-shell condition, $k^{+}\rightarrow 0$, ${\bf{k}}_{\perp}\rightarrow 0$ does not necessarily lead to soft divergences. If $k^{+}\rightarrow 0$ faster than ${\bf{k}}_{\perp}^{2}$, then $k^{-}\rightarrow \infty$. This type of divergence is a spurious IR divergence. It is a manifestation of UV divergence of the equal-time theory and hence will not be considered in this paper. The true soft divergence of light front theory is when all the components of $k^{\mu}$ (including $k^{-}$) approach $0$. In the time-ordered Hamiltonian perturbation theory, which is applied in Sec.\ref{sec:cancellation}, the IR divergences occur when the energy transferred at a vertex is zero, and thus in turn are manifested in the form of vanishing energy denominators in the perturbation series.

  \subsection{Coherent state formalism - an introduction} \label{sec:formalism introduction}
Conventionally, Fock representation is used for describing external states in a scattering process. Chung had suggested that this was an inappropriate choice because of which IR divergences appear in QED, and an alternate representation was proposed in which the transition matrix elements were shown to be free of IR divergences to all orders \cite{chung}. In this representation, an external electron is superposed with an infinite number of soft photons. Kulish and Faddeev observed that the choice of asymptotic states is associated with the form of asymptotic Hamiltonian used \cite{kulishfaddeev}. We give below a brief description of the method of asymptotic dynamics, which is used to construct the asymptotic states. The states thus obtained, when employed in the calculation of S-matrix elements, lead to IR-finite transition amplitudes.

The time-evolution operator in light front field theory is given by
\begin{equation} \label{eq:time_evoultion_op}
U(x^{+},x^{+}_{0})=T^{+}\,\text{exp}\bigg(-i\int_{x^{+}_{0}}^{x^{+}}dx'^{+}\,H_{\text{int}}(x'^{+})\bigg),
\end{equation}
where $H_{\text{int}}$ is the interaction part of the Hamiltonian and $T^{+}$ stands for the light front time-ordered product. The M\o{}ller operators are defined using the time-evolution operator as
\begin{equation} \label{eq:evol_op_general}
\begin{split}
\Omega_{\pm}\equiv \ & U(0,\mp\infty)\\
=\ &T^{+}\,\text{exp}\bigg(-i\int_{\mp\infty}^{0}dx'^{+}\,H_{\text{int}}(x'^{+})\bigg),
\end{split}
\end{equation}
and the S-matrix element is given by
\begin{equation} \label{eq:mat_el_with_H_int}
\begin{split}
\mel{f}{S}{i}=&\mel{f}{\Omega^{\dagger}_{-}\Omega_{+}}{i}\\
=&\mel{f}{T^{+}\,\text{exp}\bigg(-i\int_{-\infty}^{\infty}dx'^{+}\,H_{\text{int}}(x'^{+})\bigg)}{i}.
\end{split}
\end{equation}
The evolution of asymptotic states is conventionally presumed to be governed by the free Hamiltonian $H_{0}$ as
\begin{equation} \label{eq:free_evolution}
\ket{s(t)}=e^{-ix^{+}H_{0}}\ket{s(0)}.
\end{equation}
However, this assumption is incorrect for theories involving long-range interactions and for theories where asymptotic states are bound states. To deal with the presence of nonvanishing interactions in the large-time limit, this assumption is modified in the coherent state formalism. The evolution of asymptotic states is now considered to be governed by
\begin{equation} \label{eq:asymptotic_evolution}
\ket{s(t)}=e^{-ix^{+}H_{A(\Delta)}}\ket{s(0)},
\end{equation}
where $H_{A(\Delta)}$ is the asymptotic Hamiltonian which is obtained by taking the $|x^{+}|\rightarrow \infty$ limit of the full interaction Hamiltonian, $H_\text{int}$. Thus, $H_{A(\Delta)}$ incorporates the long-range interactions which are responsible for IR divergences. These long-range interactions describe emission/absorption of soft gauge bosons and the splitting of a massless particle into collinear massless particles. The parameter $\Delta$ appearing in $H_{A(\Delta)}$ is a cutoff for the separation of long-range interactions from $H_{\text{int}}$. The asymptotic M\o{}ller operators are defined analogously,
\begin{equation} \label{eq:evol_op_asymp}
\begin{split}
\Omega_{A(\Delta)\pm}\equiv \ & U_{A(\Delta)}(0,\mp\infty)\\
=\ &T^{+}\,\text{exp}\bigg(-i\int_{\mp\infty}^{0}dx'^{+}\,H_{\Delta}(x'^{+})\bigg).
\end{split}
\end{equation}
Now onwards, $H_{A(\Delta)}$ will be replaced by $H_{\Delta}$ for brevity.

A coherent state can now be defined as the state formed by operating the corresponding asymptotic M\o{}ller operator on a Fock state. Thus, initial (incoming) and final (outgoing) coherent states are given by
\begin{equation} \label{eq:coh_state_defn}
\begin{split}
&\ket{s:\text{coh}}\equiv \Omega_{A(\Delta)+}^{\dagger}\ket{s}\\
&\bra{s:\text{coh}}\equiv \bra{s}\Omega_{A(\Delta)-},
\end{split}
\end{equation}
respectively.

  \subsection{Construction of $\ket{q_{p_{3}}\bar{q}_{p_{4}}:\text{\normalfont{coh}}}$} \label{sec:constructing qqbar_coh}
The transition amplitude for the process $e^{+}e^{-}\rightarrow 2\; jets$, in Fock basis, is given by 
\begin{equation} \label{eq:Fock_amplitude}
T_{fi}=\mel{q_{p_{3}}\bar{q}_{p_{4}}}{T}{e^{-}_{p_{1}}e^{+}_{p_{2}}}
\end{equation}
and can be easily shown to contain IR divergences. Here, the subscript in $p_i$ denotes, in addition to the momentum of an external fermion, the other quantum numbers like color and spin also. In Sec.\ref{sec:cancellation}, we show that if a coherent state basis is used instead of the Fock basis, the IR divergences are eliminated and the modified expression for the amplitude, given by
\begin{equation} \label{eq:Tfi_2_jets}
T_{fi}=\mel{q_{p_{3}}\bar{q}_{p_{4}}:\text{coh}}{T}{e^{-}_{p_{1}}e^{+}_{p_{2}}},
\end{equation}
is IR finite. Note that since we restrict ourselves to IR divergences in QCD, the initial state is taken to be a Fock state i.e., the lowest-order term in the series expansion (in electromagnetic coupling) of the coherent state $\ket{e^{-}_{p_{1}}e^{+}_{p_{2}}:\text{coh}}$. The outgoing coherent state $\ket{q_{p_{3}}\bar{q}_{p_{4}}:\text{coh}}$, to be used in Sec.\ref{sec:cancellation} for calculating Eq.(\ref{eq:Tfi_2_jets}), is constructed to $\order{g^2}$, which is the required order in the following discussion.
Using Eq.(\ref{eq:evol_op_asymp}) and Eq.(\ref{eq:coh_state_defn}), the outgoing coherent state $\ket{q_{p_{3}}\bar{q}_{p_{4}}:\text{coh}}$ is given by
\begin{equation} \label{eq:time-ordered_coh_st}
\begin{split}
\ket{q_{p_{3}}\bar{q}_{p_{4}}:\text{coh}}=&\Omega^{\dagger}_{A(\Delta)-}\ket{q_{p3}\bar{q}_{p4}}\\
=&T^{+}\,\text{exp}\bigg(i\int_{\infty}^{0}dx'^{+}\,H_{\Delta}(x'^{+})\bigg)\ket{q_{p_{3}}\bar{q}_{p_{4}}}\\
=&T^{+}\,\text{exp}\bigg(-i\int_{0}^{\infty}dx'^{+}\,H_{\Delta}(x'^{+})\bigg)\ket{q_{p_{3}}\bar{q}_{p_{4}}}.
\end{split}
\end{equation}
The $|x^{+}|\rightarrow \infty$ limit of the interaction Hamiltonian $H_{\text{int}}$ in Eq.(\ref{eq:QCD_Hamiltonian_splitup}) leads to the asymptotic Hamiltonian $H_{\Delta}(x'^{+})$ in Eq.(\ref{eq:time-ordered_coh_st}). The interaction terms $V_{1}$ and $W_{1}$, given by Eqs.(\ref{eq:V_1}) and (\ref{eq:W_1}) respectively, are the only ones that contribute to $\ket{q_{p_{3}}\bar{q}_{p_{4}}:\text{coh}}$ in the calculation of $\mel{q_{p_{3}}\bar{q}_{p_{4}}:\text{coh}}{T}{e^{-}_{p1}e^{+}_{p2}}$ to $\mathcal{O}(g^{2})$. Hence, the coherent state up to $\mathcal{O}(g^{2})$, considering only the terms that contribute to the process under consideration, is
\begin{equation} \label{eq:qqbar_coh_st_V1_W1}
\begin{split}
\ket{q_{p_{3}}\bar{q}_{p_{4}}:\text{coh}}=& \bigg(1+(-i)\int_{0}^{\infty}dx_{1}^{+}\,V_{1}(x_{1}^{+})\Theta_{\Delta}+(-i)\int_{0}^{\infty}dx_{1}^{+}\,W_{1}(x_{1}^{+})\Theta_{\Delta}\bigg)\ket{q_{p_{3}}\bar{q}_{p_{4}}}\\
\equiv &\ket{q_{p_{3}}\bar{q}_{p_{4}}}+\Omega^{\dagger}_{\Delta(V_{1})}\ket{q_{p_{3}}\bar{q}_{p_{4}}}+\Omega^{\dagger}_{\Delta(W_{1})}\ket{q_{p_{3}}\bar{q}_{p_{4}}},
\end{split}
\end{equation}
where $\Theta_{\Delta}$ is a product of Heaviside step functions and ensures that only those terms in the potential are included that survive in the limit $|x^{+}|\rightarrow \infty$, as explained below.\\
Using Eq.(\ref{eq:V_1}) and Eq.(\ref{eq:Fourier_fields}), the second term of Eq.(\ref{eq:qqbar_coh_st_V1_W1}) becomes
\begin{equation} \label{eq:V_1_omega}
\begin{split}
& \Omega^{\dagger}_{\Delta(V_{1})}\ket{q_{p_{3}}\bar{q}_{p_{4}}}\\
=&(-i)\int_{0}^{\infty}dx_{1}^{+}\int d^{2}{\bf{x}}_{1\perp}dx^{-}_{1} g\bar{\Psi}\gamma^{\mu}T^{a}\Psi A^{a}_{\mu}\Theta_{\Delta}\ket{q_{p_{3}}\bar{q}_{p_{4}}}\\
=& (-i)g\int_{0}^{\infty}dx_{1}^{+}\int d^{2}{\bf{x}}_{1\perp}dx^{-}_{1}\int \prod_{i}^{}[dk_{i}]\Theta_{\Delta}(b^{\dagger}_{k_{1}}\bar{u}_{k_{1}}e^{ik_{1}{\cdot}x_{1}}+d_{k_{1}}\bar{v}_{k_{1}}e^{-ik_{1}{\cdot}x_{1}})\\
& \gamma^{\mu}T^{a}(b_{k_{2}}u_{k_{2}}e^{-ik_{2}{\cdot}x_{1}}+d^{\dagger}_{k_{2}}v_{k_{2}}e^{ik_{2}{\cdot}x_{1}})(a_{k_{3}}\epsilon^{a}_{\mu k_{3}}e^{-ik_{3}{\cdot}x_{1}}+a^{\dagger}_{k_{3}}\epsilon^{*a}_{\mu k_{3}}e^{ik_{3}{\cdot}x_{1}})\ket{q_{p_{3}}\bar{q}_{p_{4}}},
\end{split}
\end{equation}
where the symbol $\prod$ stands for product over momenta $k_{i}$ and summation over the spin and color quantum numbers. The presence of the oscillating factors $\int dx^{+} e^{i\sum_{i}k^{-}_{i}x^{+}}$ in Eq.(\ref{eq:V_1_omega}) implies that only those terms survive in the $|x^{+}|\rightarrow \infty$ limit, for which the condition $\sum_{i}k^{-}_{i}\rightarrow 0$ is satisfied. Thus, in the limit $|x^{+}|\rightarrow \infty$,
\begin{equation}
\begin{split}
& \Omega^{\dagger}_{\Delta(V_{1})}\ket{q_{p_{3}}\bar{q}_{p_{4}}}\\
=& (-i)g \int_{0}^{\infty}dx_{1}^{+}\int d^{2}{\bf{x}}_{1\perp}dx^{-}_{1}\int \prod_{i}^{}[dk_{i}]\Theta_{\Delta}(\bar{u}_{k_{1}}\gamma^{\mu}T^{a}u_{k_{2}}\epsilon^{*a}_{\mu k_{3}}b^{\dagger}_{k_{1}}b_{k_{2}}a^{\dagger}_{k_{3}}\\
& e^{i(k_{1}-k_{2}+k_{3}){\cdot}x_{1}}+\bar{v}_{k_{1}}\gamma^{\mu}T^{a}v_{k_{2}}\epsilon^{*a}_{\mu k_{3}}d_{k_{1}}d^{\dagger}_{k_{2}}a^{\dagger}_{k_{3}}e^{-i(k_{1}-k_{2}-k_{3}){\cdot}x_{1}})b^{\dagger}_{p_{3}}d^{\dagger}_{p_{4}}\ket{0}.
\end{split}
\end{equation}
The rest of the terms in Eq.(\ref{eq:V_1_omega}) are either zero (viz. $...a_{k_{3}}\ket{q_{p_{3}}\bar{q}_{p_{4}}}=0$) or eliminated by $\Theta_{\Delta}$ function because $\sum_{i}k^{-}_{i}$ can never approach zero for those terms. Performing space and time integrations and using
\begin{equation}
\begin{split}
& b^{\dagger}_{k_{1}}b_{k_{2}}a^{\dagger}_{k_{3}}b^{\dagger}_{p_{3}}d^{\dagger}_{p_{4}}\ket{0}=\delta({k_{2}}-p_{3})a^{\dagger}_{k_{3}}b^{\dagger}_{k_{1}}d^{\dagger}_{p_{4}}\ket{0}\\
& d_{k_{1}}d^{\dagger}_{k_{2}}a^{\dagger}_{k_{3}}b^{\dagger}_{p_{3}}d^{\dagger}_{p_{4}}\ket{0}=-\delta({k_{1}}-p_{4})a^{\dagger}_{k_{3}}b^{\dagger}_{p_{3}}d^{\dagger}_{k_{2}}\ket{0},
\end{split}
\end{equation}
where $\delta(k-p)\equiv \delta^{2}({\bf{k}}_{\perp}-{\bf{p}}_{\perp})\delta(k^{+}-p^{+})\delta_{spins}\delta_{colors}\delta_{flavors}$, to integrate over $k_{1}$ and $k_{2}$, one obtains
\begin{equation} \label{eq:qqbar_coh_st_V_1}
\begin{split}
\Omega^{\dagger}_{\Delta(V_{1})}\ket{q_{p_{3}}\bar{q}_{p_{4}}}= -g\int [dk]\Theta_{\Delta}& \Bigg[\frac{\bar{u}_{(p_{3}-k)}\slashed{\epsilon}^{*a}_{k}T^{a}u_{p_{3}}a^{\dagger}_{k}b^{\dagger}_{(p_{3}-k)}d^{\dagger}_{p_{4}}\ket{0}}{\sqrt{2p^{+}_{3}}\sqrt{2(p^{+}_{3}-k^{+})}(p^{-}_{3}-k^{-}-(p_{3}-k)^{-})}\\
& +\frac{\bar{v}_{p_{4}}\slashed{\epsilon}^{*a}_{k}T^{a}v_{(p_{4}-k)}a^{\dagger}_{k}b^{\dagger}_{p_{3}}d^{\dagger}_{(p_{4}-k)}\ket{0}}{\sqrt{2p^{+}_{4}}\sqrt{2(p^{+}_{4}-k^{+})}(p^{-}_{4}-k^{-}-(p_{4}-k)^{-})}\Bigg].
\end{split}
\end{equation}
A similar procedure leads to
\begin{equation} \label{eq:qqbar_coh_st_W_1}
\begin{split}
& \Omega^{\dagger}_{\Delta(W_{1})}\ket{q_{p_{3}}\bar{q}_{p_{4}}}\\
=& g^{2}\int [dk]\Theta_{\Delta}\Bigg[\frac{1}{(2\pi)^{3/2}\sqrt{2p^{+}_{3}}\sqrt{2p^{+}_{4}}\sqrt{2(p^{+}_{3}+p^{+}_{4}-k^{+})}}\\
& \frac{\bar{u}_{k}\gamma^{+}T^{a}u_{p_{3}}\bar{v}_{p_{4}}\gamma^{+}T^{a}v_{(p_{3}+p_{4}-k)}b^{\dagger}_{k}d^{\dagger}_{(p_{3}+p_{4}-k)}\ket{0}}{(p^{+}_{3}-k^{+})^{2}(p^{-}_{3}+p^{-}_{4}-k^{-}-(p_{3}+p_{4}-k)^{-})}\Bigg]
\end{split}
\end{equation}
for the third term in Eq.(\ref{eq:qqbar_coh_st_V1_W1}).
Note that $\Omega^{\dagger}_{\Delta(V_{1})}\ket{q_{p_{3}}\bar{q}_{p_{4}}}$ is an $\mathcal{O}(g)$ and $\Omega^{\dagger}_{\Delta(W_{1})}\ket{q_{p_{3}}\bar{q}_{p_{4}}}$ is an $\mathcal{O}(g^{2})$ contribution to the outgoing coherent state.

  \subsection{The asymptotic region} \label{sec:asymptotic_region}
\begin{figure}[ht]
\centering
{\includegraphics[width=.3\columnwidth]{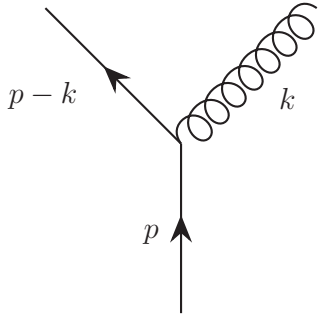}}
\caption{The qqg vertex}
\label{fig:qqg_vertex}
\end{figure}

In Sec.\ref{sec:constructing qqbar_coh}, the function $\Theta_{\Delta}$ is introduced to ensure that only those terms appear in the definition of coherent state for which the condition $\sum_{i}k^{-}_{i}< \Delta$ is satisfied. This leads to the vanishing energy denominators in Eqs.(\ref{eq:qqbar_coh_st_V_1}) and (\ref{eq:qqbar_coh_st_W_1}). Thus, $\Theta_{\Delta}$ restricts the region of phase space to only those values where the energy transferred at a vertex tends to zero in the coherent state. In this section, we illustrate this idea of the region of phase space to be included in asymptotic states through the example of a $qqg$ vertex. Consider the quark-quark-gluon vertex shown in Fig.\ref{fig:qqg_vertex}.

Without loss of generality, the momentum of incoming quark can be written as

\begin{equation}
p^{\mu}=\bigg(p^{+}, \frac{p^{2}_{x}}{2p^{+}}, p_{x}, 0\bigg).
\end{equation}
The momentum of outgoing gluon is parametrized as 

\begin{equation}
k^{\mu}=\bigg(\alpha p^{+}, \frac{(\alpha+\beta)^{2}p^{2}_{x}}{2\alpha p^{+}}, (\alpha+\beta)p_{x}, 0\bigg),
\end{equation}
where $0\leq\alpha\leq 1$ and $\beta$ can be either positive or negative. Applying 3-momentum conservation, the momentum of outgoing quark becomes 

\begin{equation}
(p-k)^{\mu}=\bigg((1-\alpha)p^{+}, \frac{(1-\alpha-\beta)^{2}p^{2}_{x}}{2(1-\alpha)p^{+}}, (1-\alpha-\beta)p_{x}, 0\bigg).
\end{equation}
In the Hamiltonian formulation, IR divergences occur when the light front energy transferred at the vertex tends to zero. Hence, for IR divergences to occur in this example, the criterion is $|p^{-}-k^{-}-(p-k)^{-}|\rightarrow 0$ which can be implemented through the Heaviside step function

\begin{equation}
\Theta_{\Delta}=\Theta(\Delta-|p^{-}-k^{-}-(p-k)^{-}|).
\end{equation}
In terms of the parameters $\alpha$ and $\beta$,

\begin{equation}
\begin{split}
\Theta_{\Delta}(\alpha,\beta)=& \Theta\bigg(\Delta-\frac{p^{2}_{x}}{2p^{+}}\bigg|1-\frac{(\alpha+\beta)^{2}}{\alpha}-\frac{(1-\alpha-\beta)^{2}}{(1-\alpha)}\bigg|\bigg)\\
=& \Theta\bigg(\Delta-\frac{p^{2}_{x}}{2p^{+}}\frac{\beta^{2}}{\alpha(1-\alpha)}\bigg).
\end{split}
\end{equation}
Thus, infrared divergences arise due to following possibilities:

{\it Case 1:} Purely (hard) collinear gluon\\
$\beta\rightarrow 0,\ \alpha\neq 0$\\
$\implies \Theta_{\Delta}=\theta(\alpha-\Delta)\theta(\Delta-\beta)$.\\

{\it Case 2:} Soft + Soft-collinear gluon\\
$\beta\rightarrow 0,\ \alpha\rightarrow 0$ such that $k^{-}\rightarrow 0$\\
$\implies \Theta_{\Delta}=\theta(\Delta-\alpha)[\theta(\sqrt{\alpha\Delta}-\beta)-\theta(-\sqrt{\alpha\Delta}-\beta)]$.\\
In addition to these two cases of `true' IR divergences, there also exists spurious IR divergence in light front theories which occurs when the parameters $\beta\rightarrow 0,\ \alpha\rightarrow 0$, but the gluon energy component $k^{-}\rightarrow\infty$. The spurious IR divergence is a manifestation of ultraviolet divergence of the equal-time theory.

The constraint $\Theta_{\Delta}=\Theta(\Delta-|p^{-}-k^{-}-(p-k)^{-}|)$ includes both the soft and collinear regions and hence in this formalism, both soft and collinear divergences can be shown to cancel at the amplitude level in inclusive cross sections like in the process under consideration.

\section{Cancellation of IR divergences to $\mathcal{O}(g^{2})$}\label{sec:cancellation}
The transition amplitude for the process $e^{+}e^{-}\rightarrow 2\; \text{jets}$ in coherent state basis is given by
\begin{equation}
T_{fi}=\mel{q_{p_{3}}\bar{q}_{p_{4}}:\text{coh}}{T}{e^{-}_{p_{1}}e^{+}_{p_{2}}}
\end{equation}
Employing the expansion for the outgoing coherent state $\ket{q_{p_{3}}\bar{q}_{p_{4}}:\text{coh}}$ upto $\mathcal{O}(g^{2})$ given by Eq.(\ref{eq:qqbar_coh_st_V1_W1})
\begin{equation*}
\ket{q_{p_{3}}\bar{q}_{p_{4}}:\text{coh}}=\ket{q_{p_{3}}\bar{q}_{p_{4}}}+\Omega^{\dagger}_{\Delta(V_{1})}\ket{q_{p_{3}}\bar{q}_{p_{4}}}+\Omega^{\dagger}_{\Delta(W_{1})}\ket{q_{p_{3}}\bar{q}_{p_{4}}},
\end{equation*}
the transition amplitude to $\mathcal{O}(g^{2})$ becomes
\begin{equation}
\begin{split}
T_{fi}=& \mel{q_{p_{3}}\bar{q}_{p_{4}}:\text{coh}}{T}{e^{-}_{p_{1}}e^{+}_{p_{2}}}\\
=& \mel{q_{p_{3}}\bar{q}_{p_{4}}}{T}{e^{-}_{p_{1}}e^{+}_{p_{2}}}\\
& +\mel{q_{p_{3}}\bar{q}_{p_{4}}}{\Omega_{\Delta(V_{1})}T}{e^{-}_{p_{1}}e^{+}_{p_{2}}}\\
& +\mel{q_{p_{3}}\bar{q}_{p_{4}}}{\Omega_{\Delta(W_{1})}T}{e^{-}_{p_{1}}e^{+}_{p_{2}}}.
\end{split}
\end{equation}
The initial state $\ket{e^{+}e^{-}}$ is considered as a Fock state, and not a coherent state, since only QCD IR divergences are being dealt with here. Thus, the time-ordered perturbation series for the transition amplitude is
\begin{equation} \label{eq:pert_expn_full}
\begin{split}
T_{fi}=& \mel{q_{p_{3}}\bar{q}_{p_{4}}:\text{coh}}{T}{e^{-}_{p_{1}}e^{+}_{p_{2}}}\\
=& \mel{q_{p_{3}}\bar{q}_{p_{4}}}{V+V\frac{1}{p^{-}_{i}-H_{0}}V+V\frac{1}{p^{-}_{i}-H_{0}}V\frac{1}{p^{-}_{i}-H_{0}}V+...}{e^{-}_{p_{1}}e^{+}_{p_{2}}}\\
& +\mel{q_{p_{3}}\bar{q}_{p_{4}}}{\Omega_{\Delta(V_{1})}V+V\frac{1}{p^{-}_{i}-H_{0}}V+V\frac{1}{p^{-}_{i}-H_{0}}V\frac{1}{p^{-}_{i}-H_{0}}V+...}{e^{-}_{p_{1}}e^{+}_{p_{2}}}\\
& +\mel{q_{p_{3}}\bar{q}_{p_{4}}}{\Omega_{\Delta(W_{1})}V+V\frac{1}{p^{-}_{i}-H_{0}}V+V\frac{1}{p^{-}_{i}-H_{0}}V\frac{1}{p^{-}_{i}-H_{0}}V+...}{e^{-}_{p_{1}}e^{+}_{p_{2}}}\\
& +...,
\end{split}
\end{equation}
where $V$ denotes a generic interaction potential, $p^{-}_{i}$ is the total energy of the initial state and $H_{0}$ is the free light front Hamiltonian. Note that at $\mathcal{O}(g^{0})$, Eq.(\ref{eq:pert_expn_full}) leads to the tree-level QED diagram with an outgoing $\ket{q\bar{q}}$ Fock state and hence contains no QCD IR divergences. Terms of odd order in $g$ are absent in Eq.(\ref{eq:pert_expn_full}). Also, the initial state $\ket{e^{-}e^{+}}$ connects to a $\ket{q\bar{q}}$ state only through a photon. Hence, the lowest order at which QCD IR divergences occur is $\mathcal{O}(g^{2})$, in which Eq.(\ref{eq:pert_expn_full}) reduces to
\begin{equation} \label{eq:pert_expn_reqd}
\begin{split}
T_{fi}=& \mel{q_{p_{3}}\bar{q}_{p_{4}}}{V_{1}\frac{1}{p^{-}_{i}-H_{0}}V_{1}\frac{1}{p^{-}_{i}-H_{0}}V_{em}\frac{1}{p^{-}_{i}-H_{0}}V_{em}}{e^{-}_{p_{1}}e^{+}_{p_{2}}}\\
& +\mel{q_{p_{3}}\bar{q}_{p_{4}}}{W_{1}\frac{1}{p^{-}_{i}-H_{0}}V_{em}\frac{1}{p^{-}_{i}-H_{0}}V_{em}}{e^{-}_{p_{1}}e^{+}_{p_{2}}}\\
& +\mel{q_{p_{3}}\bar{q}_{p_{4}}}{\Omega_{\Delta(V_{1})}V_{1}\frac{1}{p^{-}_{i}-H_{0}}V_{em}\frac{1}{p^{-}_{i}-H_{0}}V_{em}}{e^{-}_{p_{1}}e^{+}_{p_{2}}}\\
& +\mel{q_{p_{3}}\bar{q}_{p_{4}}}{\Omega_{\Delta(W_{1})}V_{em}\frac{1}{p^{-}_{i}-H_{0}}V_{em}}{e^{-}_{p_{1}}e^{+}_{p_{2}}},
\end{split}
\end{equation}
where $V_{em}$ is the electromagnetic interaction. The first two terms in Eq.(\ref{eq:pert_expn_reqd}) would be the contributions obtained if one used the Fock basis instead of coherent state basis. The additional two terms are present because coherent state basis is used to calculate the transition amplitude. In the rest of this section, both these contributions are calculated. It is shown that each contribution contains IR-divergent expressions and that the IR divergences cancel between them.

  \subsection{$\mathcal{O}(g^{0})$ coherent state contribution to $\mathcal{O}(g^{2})$ transition amplitude}\label{sec:fock_contri}
\begin{figure}[ht]
\centering
\subfloat[$T_{1a}$]
{\includegraphics[width=0.38\columnwidth]{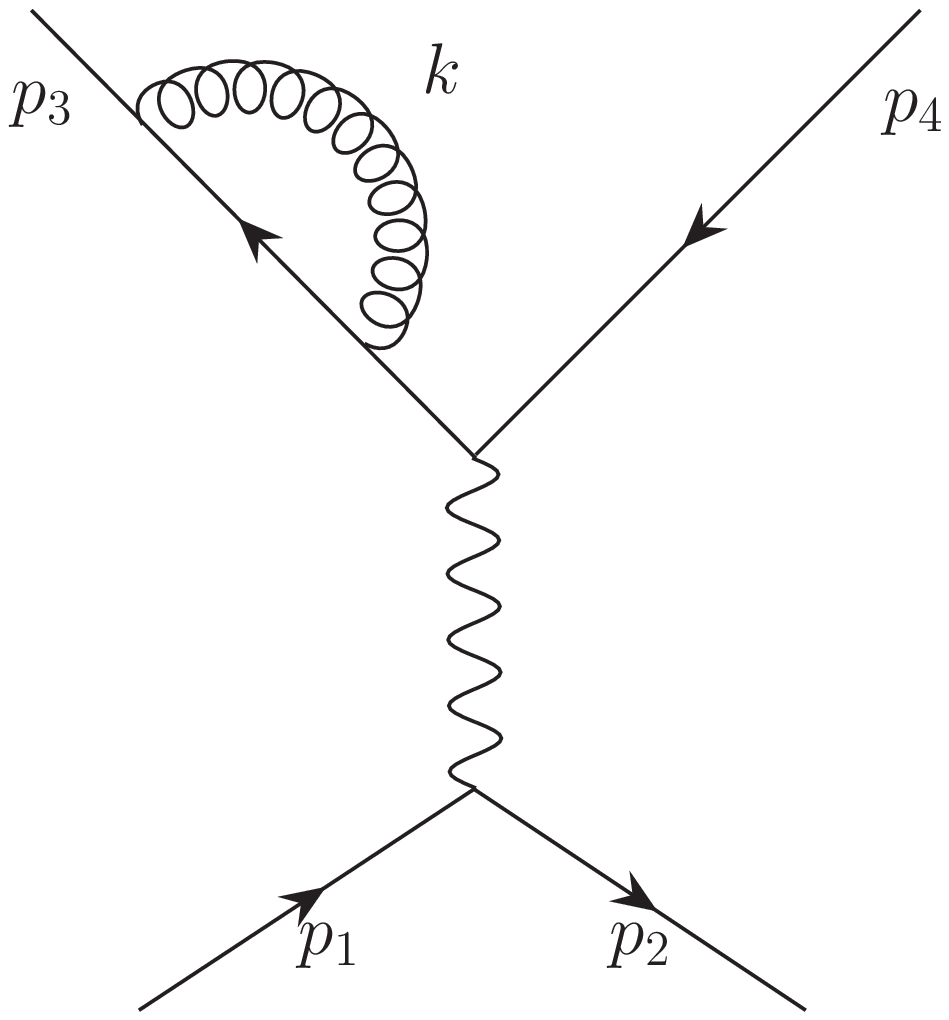}}
\hspace{.75cm}
\subfloat[$T_{1b}$]
{\includegraphics[width=0.38\columnwidth]{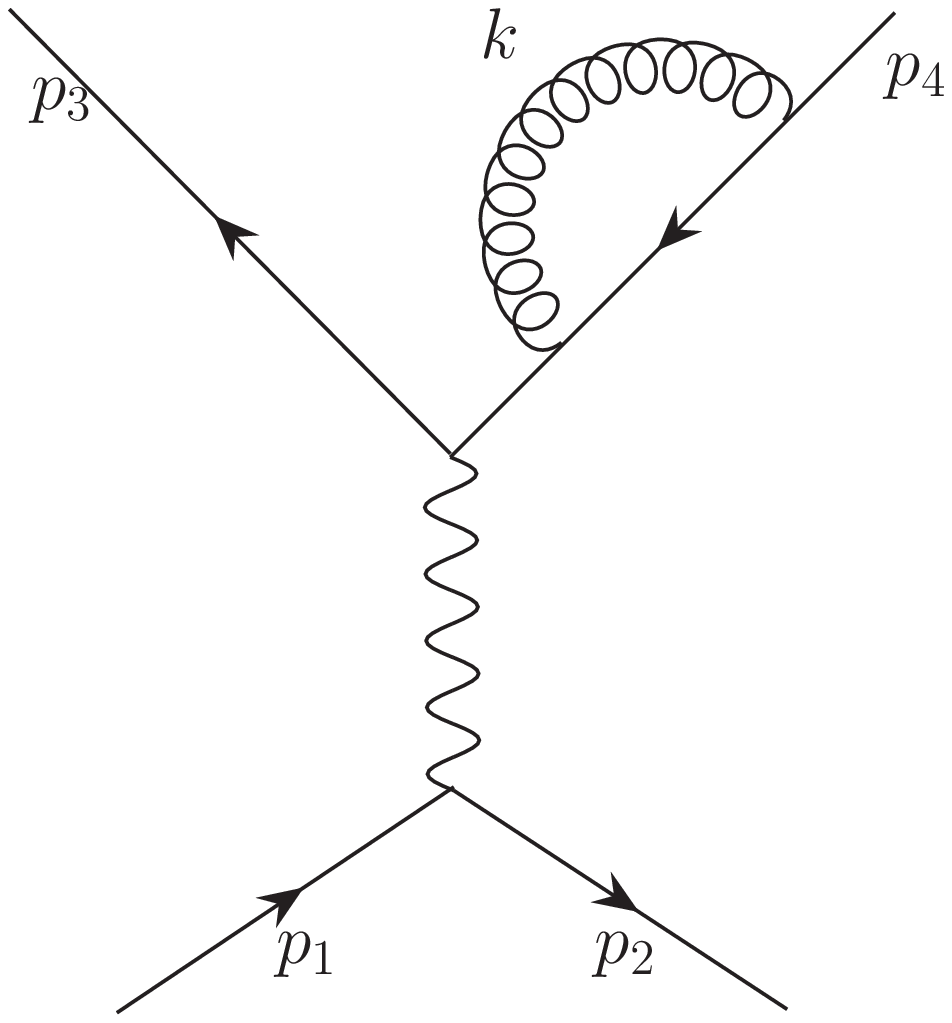}}
\hspace{.75cm}
\subfloat[$T_{1c}$]
{\includegraphics[width=0.38\columnwidth]{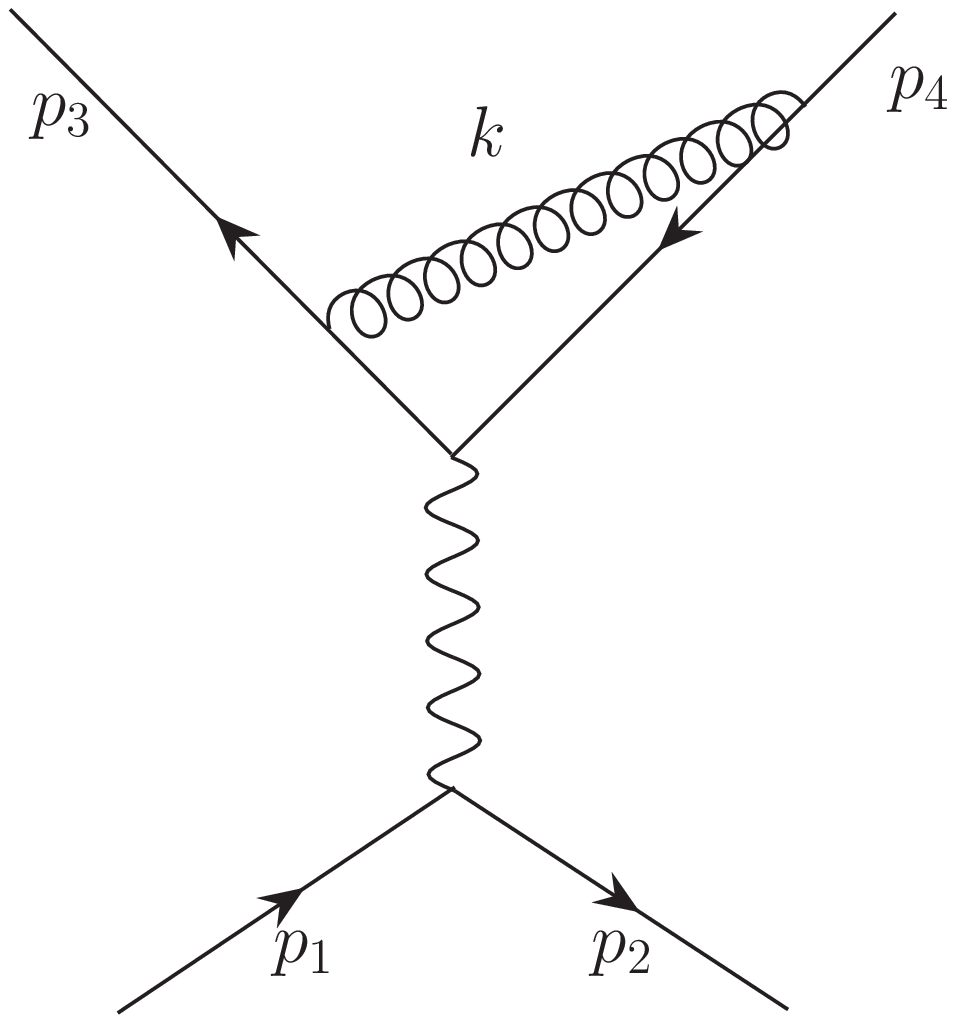}}
\hspace{.75cm}
\subfloat[$T_{1d}$]
{\includegraphics[width=0.38\columnwidth]{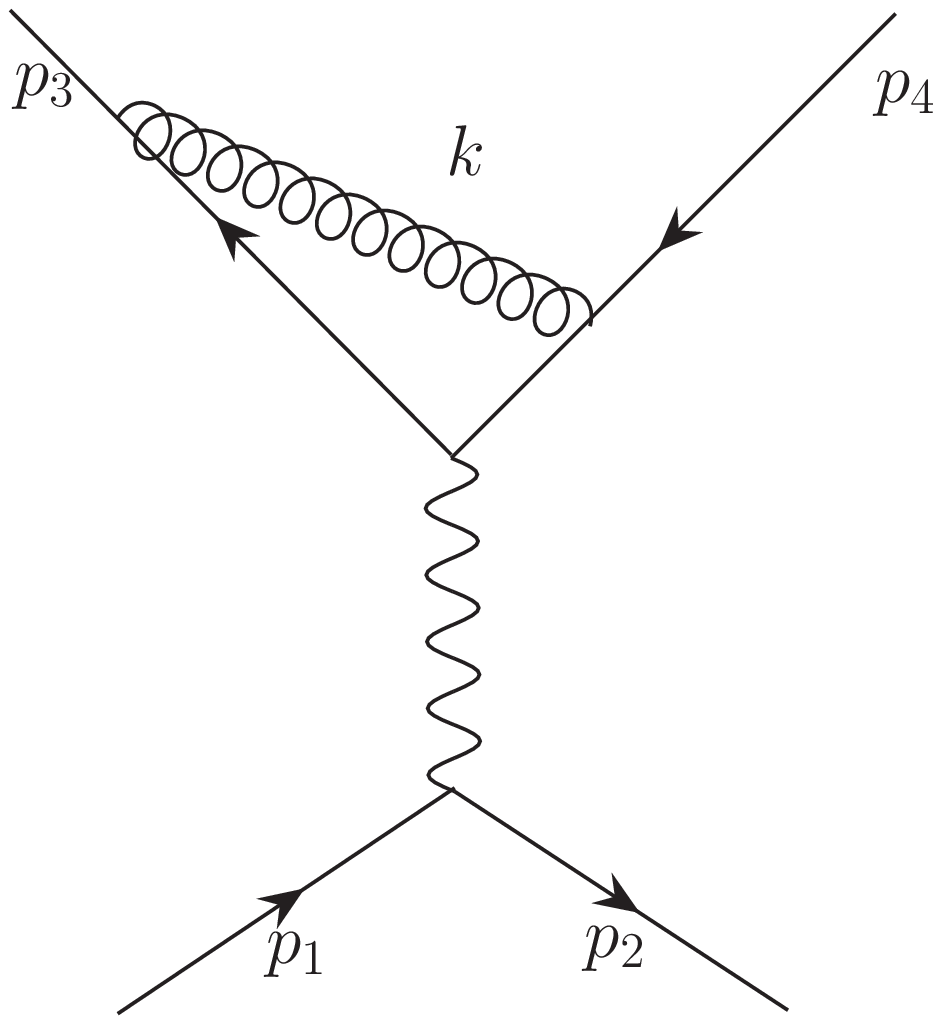}}
\caption{Regular diagrams corresponding to outgoing Fock state}
\label{fig:regular_fock_diagrams}
\end{figure}

\begin{figure}[ht]
\centering
{\includegraphics[width=0.38\columnwidth]{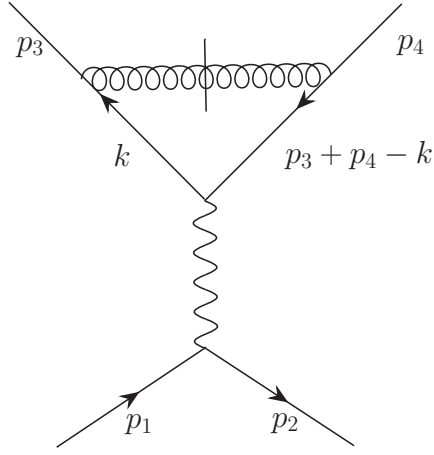}}
\caption{Instantaneous diagram corresponding to outgoing Fock state}
\label{fig:inst_fock_diagram}
\end{figure}

The first two terms of Eq.(\ref{eq:pert_expn_reqd}) represent the contribution of $\mathcal{O}(g^{0})$ term of $\ket{q_{p_{3}}\bar{q}_{p_{4}}:\text{coh}}$ to $T_{fi}$ ($\mathcal{O}(g^{2})$) in the coherent state basis. These terms are also the ones that would give the $\mathcal{O}(g^{2})$ transition amplitude in the Fock basis. The expressions for these, which are obtained by inserting complete set of states between operators and evaluating the resulting inner products, are given below. To illustrate the technique, one of the terms of Eq.(\ref{eq:pert_expn_reqd}) is explicitly calculated in Appendix \ref{app:Amplitude_calculation}. The diagrams corresponding to the Fock state contribution are shown in Fig.\ref{fig:regular_fock_diagrams} and Fig.\ref{fig:inst_fock_diagram}.
We define
\begin{equation} \label{eq:T_1}
\begin{split}
T_{1}\equiv &\mel{q_{p_{3}}\bar{q}_{p_{4}}}{V_{1}\frac{1}{p^{-}_{i}-H_{0}}V_{1}\frac{1}{p^{-}_{i}-H_{0}}V_{em}\frac{1}{p^{-}_{i}-H_{0}}V_{em}}{e^{-}_{p_{1}}e^{+}_{p_{2}}}\\
=&\ T_{1a}+T_{1b}+T_{1c}+T_{1d},
\end{split}
\end{equation}
where
\begin{equation} \label{eq:T_1a}
\begin{split}
T_{1a}=& \frac{e^{2}g^{2}}{(2\pi)^{6}}\int\frac{d^{2}{\bf{k}}_{\perp}dk^{+}}{\prod_{i}^{}\sqrt{2p^{+}_{i}}}\Bigg[\frac{\bar{u}_{p_{3}}\slashed{\epsilon}^{a}_{k}u_{(p_{3}-k)}\bar{u}_{(p_{3}-k)}\slashed{\epsilon}^{*a}_{k}u_{p_{3}}\bar{u}_{p_{3}}}{2(p^{+}_{1}+p^{+}_{2})2k^{+}2p^{+}_{3}2(p^{+}_{3}-k^{+})}\\
& \frac{\slashed{\epsilon}_{(p_{1}+p_{2})}v_{p_{4}}\bar{v}_{p_{2}}\slashed{\epsilon}^{*}_{(p_{1}+p_{2})}u_{p_{1}}C_{F}\delta_{c_{3}c_{4}}}{(p^{-}_{3}-k^{-}-(p_{3}-k)^{-})(p^{-}_{1}+p^{-}_{2}-p^{-}_{3}-p^{-}_{4})(p^{-}_{1}+p^{-}_{2}-(p_{1}+p_{2})^{-})}\Bigg]
\end{split}
\end{equation}
\begin{equation} \label{eq:T_1b}
\begin{split}
T_{1b}=& \frac{e^{2}g^{2}}{(2\pi)^{6}}\int\frac{d^{2}{\bf{k}}_{\perp}dk^{+}}{\prod_{i}^{}\sqrt{2p^{+}_{i}}}\Bigg[\frac{\bar{u}_{p_{3}}\slashed{\epsilon}_{(p_{1}+p_{2})}v_{p_{4}}\bar{v}_{p_{4}}\slashed{\epsilon}^{*a}_{k}v_{(p_{4}-k)}\bar{v}_{(p_{4}-k)}\slashed{\epsilon}^{a}_{k}}{2(p^{+}_{1}+p^{+}_{2})2k^{+}2p^{+}_{4}2(p^{+}_{4}-k^{+})}\\
& \frac{v_{p_{4}}\bar{v}_{p_{2}}\slashed{\epsilon}^{*}_{(p_{1}+p_{2})}u_{p_{1}}C_{F}\delta_{c_{3}c_{4}}}{(p^{-}_{4}-k^{-}-(p_{4}-k)^{-})(p^{-}_{1}+p^{-}_{2}-p^{-}_{3}-p^{-}_{4})(p^{-}_{1}+p^{-}_{2}-(p_{1}+p_{2})^{-})}\Bigg]
\end{split}
\end{equation}
\begin{equation} \label{eq:T_1c}
\begin{split}
T_{1c}=& -\frac{e^{2}g^{2}}{(2\pi)^{6}}\int\frac{d^{2}{\bf{k}}_{\perp}dk^{+}}{\prod_{i}^{}\sqrt{2p^{+}_{i}}}\Bigg[\frac{\bar{u}_{p_{3}}\slashed{\epsilon}^{*a}_{k}u_{(p_{3}+k)}\bar{u}_{(p_{3}+k)}\slashed{\epsilon}_{(p_{1}+p_{2})}v_{(p_{4}-k)}\bar{v}_{(p_{4}-k)}}{2(p^{+}_{1}+p^{+}_{2})2k^{+}2(p^{+}_{3}+k^{+})2(p^{+}_{4}-k^{+})}\\
& \frac{\slashed{\epsilon}^{a}_{k}v_{p_{4}}\bar{v}_{p_{2}}\slashed{\epsilon}^{*}_{(p_{1}+p_{2})}u_{p_{1}}C_{F}\delta_{c_{3}c_{4}}}{(p^{-}_{4}-k^{-}-(p_{4}-k)^{-})(p^{-}_{1}+p^{-}_{2}-(p_{3}+k)^{-}-(p_{4}-k)^{-})(p^{-}_{1}+p^{-}_{2}-(p_{1}+p_{2})^{-})}\Bigg]
\end{split}
\end{equation}
\begin{equation} \label{eq:T_1d}
\begin{split}
T_{1d}=& -\frac{e^{2}g^{2}}{(2\pi)^{6}}\int\frac{d^{2}{\bf{k}}_{\perp}dk^{+}}{\prod_{i}^{}\sqrt{2p^{+}_{i}}}\Bigg[\frac{\bar{u}_{p_{3}}\slashed{\epsilon}^{a}_{k}u_{(p_{3}-k)}\bar{u}_{(p_{3}-k)}\slashed{\epsilon}_{(p_{1}+p_{2})}v_{(p_{4}+k)}\bar{v}_{(p_{4}+k)}}{2(p^{+}_{1}+p^{+}_{2})2k^{+}2(p^{+}_{3}-k^{+})2(p^{+}_{4}+k^{+})}\\
& \frac{\slashed{\epsilon}^{*a}_{k}v_{p_{4}}\bar{v}_{p_{2}}\slashed{\epsilon}^{*}_{(p_{1}+p_{2})}u_{p_{1}}C_{F}\delta_{c_{3}c_{4}}}{(p^{-}_{3}-k^{-}-(p_{3}-k)^{-})(p^{-}_{1}+p^{-}_{2}-(p_{3}-k)^{-}-(p_{4}+k)^{-})(p^{-}_{1}+p^{-}_{2}-(p_{1}+p_{2})^{-})}\Bigg].
\end{split}
\end{equation}
$T_{1a}, T_{1b}, T_{1c} $ and $T_{1d}$ correspond to the diagrams in Figs.\ref{fig:regular_fock_diagrams}(a)-\ref{fig:regular_fock_diagrams}(d), respectively.
The expressions for $T_{1a}$ through $T_{1d}$ i.e., Eqs.(\ref{eq:T_1a})-(\ref{eq:T_1d}) are IR-divergent in the following limits:\\
$T_{1a}$ : $(p^{-}_{3}-k^{-}-(p_{3}-k)^{-})\rightarrow 0$\\
$T_{1b}$ : $(p^{-}_{4}-k^{-}-(p_{4}-k)^{-})\rightarrow 0$\\
$T_{1c}$ : $(p^{-}_{4}-k^{-}-(p_{4}-k)^{-})\rightarrow 0$; $(p^{-}_{1}+p^{-}_{2}-(p_{3}+k)^{-}-(p_{4}-k)^{-})\rightarrow 0$\\
$T_{1d}$ : $(p^{-}_{3}-k^{-}-(p_{3}-k)^{-})\rightarrow 0$; $(p^{-}_{1}+p^{-}_{2}-(p_{4}+k)^{-}-(p_{3}-k)^{-})\rightarrow 0$.

Next, we define
\begin{equation} \label{eq:T_2}
\begin{split}
T_{2}\equiv &\mel{q_{p_{3}}\bar{q}_{p_{4}}}{W_{1}\frac{1}{p^{-}_{i}-H_{0}}V_{em}\frac{1}{p^{-}_{i}-H_{0}}V_{em}}{e^{-}_{p_{1}}e^{+}_{p_{2}}}\\
=& -\frac{e^{2}g^{2}}{(2\pi)^{6}}\int\frac{d^{2}{\bf{k}}_{\perp}dk^{+}}{\prod_{i}^{}\sqrt{2p^{+}_{i}}}\Bigg[\frac{\bar{u}_{p_{3}}\gamma^{+}u_{k}\bar{u}_{k}\slashed{\epsilon}_{(p_{1}+p_{2})}v_{(p_{3}+p_{4}-k)}\bar{v}_{(p_{3}+p_{4}-k)}}{2(p^{+}_{1}+p^{+}_{2})2k^{+}2(p^{+}_{3}+p^{+}_{4}-k^{+})(p^{+}_{3}-k^{+})^{2}}\\
& \frac{\gamma^{+}v_{p_{4}}\bar{v}_{p_{2}}\slashed{\epsilon}^{*}_{(p_{1}+p_{2})}u_{p_{1}}C_{F}\delta_{c_{3}c_{4}}}{(p^{-}_{3}+p^{-}_{4}-k^{-}-(p_{3}+p_{4}-k)^{-})(p^{-}_{1}+p^{-}_{2}-(p_{1}+p_{2})^{-})}\Bigg]
\end{split}
\end{equation}
which is the amplitude of the diagram in Fig.\ref{fig:inst_fock_diagram}.
$T_{2}$ given in Eq.(\ref{eq:T_2}) is IR-divergent when $(p^{-}_{3}+p^{-}_{4}-k^{-}-(p_{3}+p_{4}-k)^{-})\rightarrow 0$. Note that $T_{1}$ and $T_{2}$ calculated in this section, and hence Figs.\ref{fig:regular_fock_diagrams} and \ref{fig:inst_fock_diagram}, would have been the only contributions to the transition amplitude if Fock basis was used, thus leading to an IR-divergent transition amplitude in the Fock basis. Additional contributions arising due to the use of coherent state basis are calculated below. 

  \subsection{$\mathcal{O}(g)$ and $\mathcal{O}(g^{2})$ coherent state contributions to $\mathcal{O}(g^{2})$ transition amplitude}\label{sec:coherent_contri}
\begin{figure}[ht]
\centering
\subfloat[$T_{3a}$]
{\includegraphics[width=0.38\columnwidth]{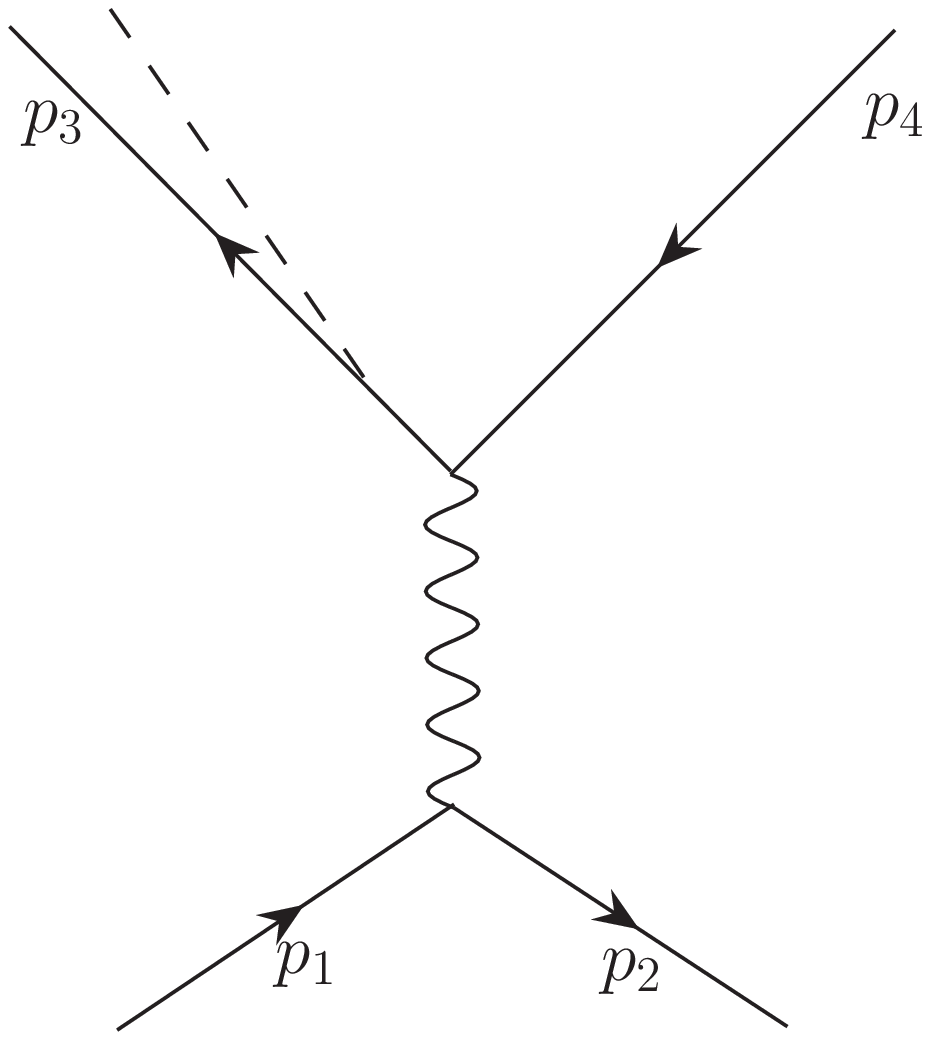}}
\hspace{.75cm}
\subfloat[$T_{3b}$]
{\includegraphics[width=0.38\columnwidth]{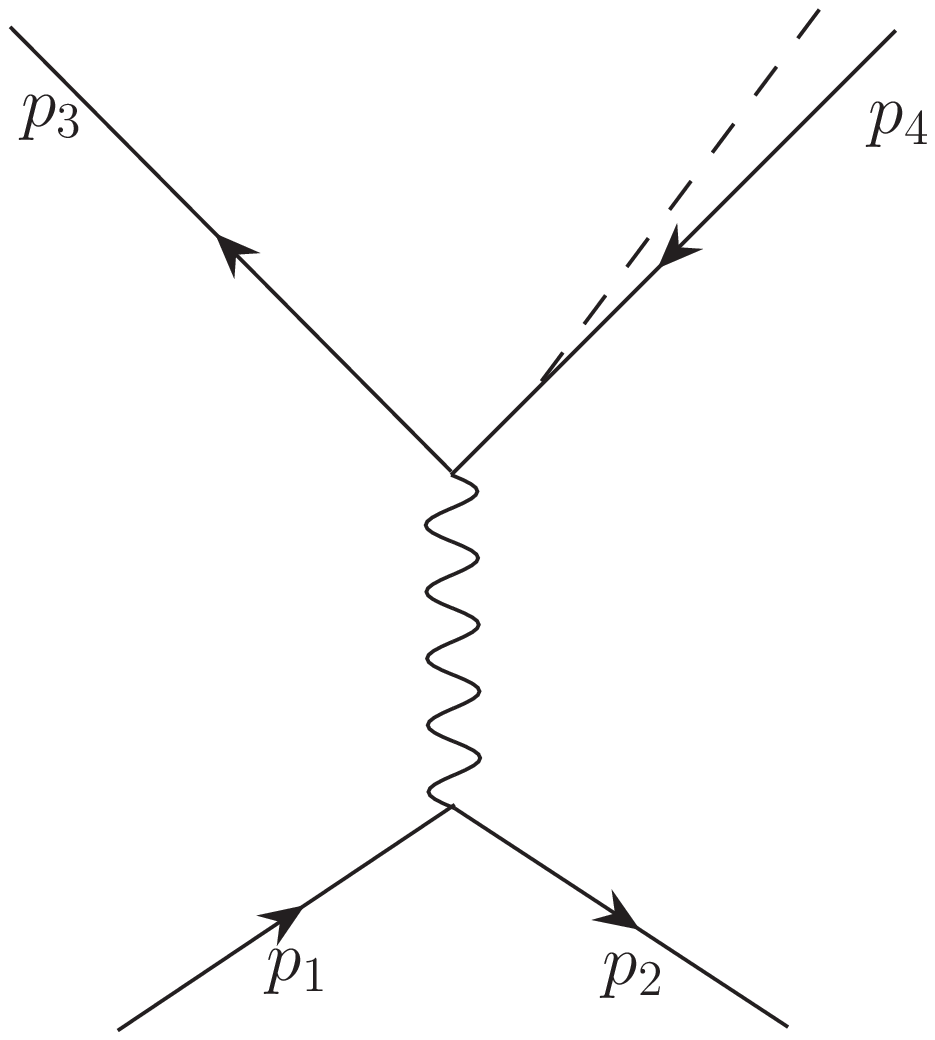}}
\hspace{.75cm}
\subfloat[$T_{3c}$]
{\includegraphics[width=0.38\columnwidth]{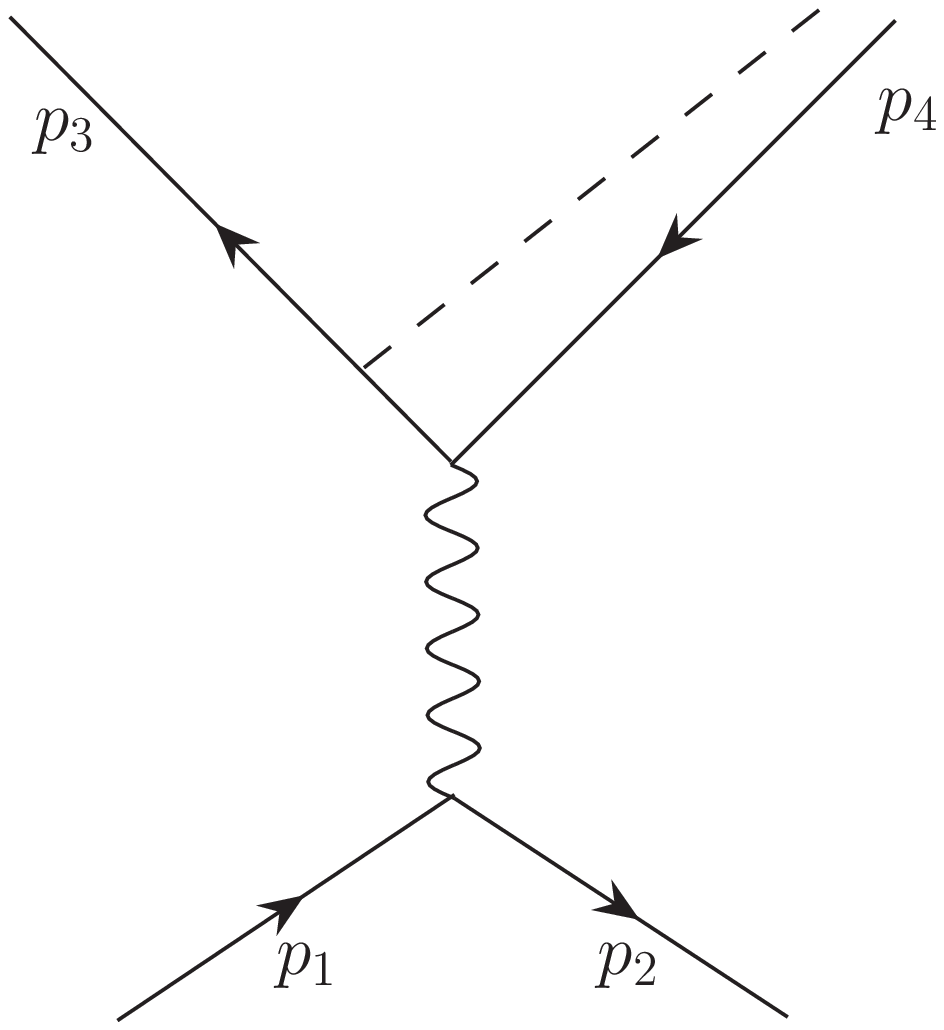}}
\hspace{.75cm}
\subfloat[$T_{3d}$]
{\includegraphics[width=0.38\columnwidth]{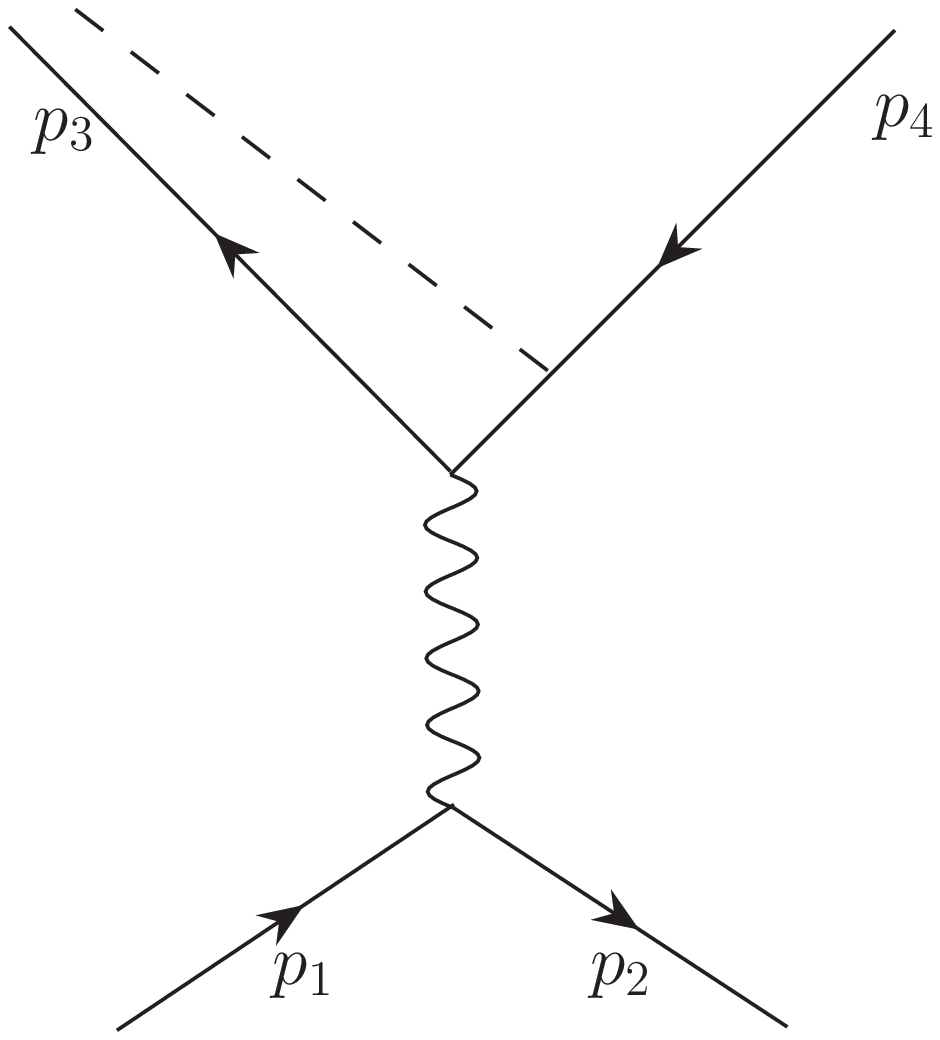}}
\caption{Regular diagrams corresponding to outgoing coherent state}
\label{fig:regular_coh_diagrams}
\end{figure}

In coherent state basis, in addition to the contribution calculated in Sec.\ref{sec:fock_contri}, there are additional contributions to the transition amplitude arising from $\mathcal{O}(g)$ and $\mathcal{O}(g^2)$ terms in the expansion of coherent state. The $\mathcal{O}(g)$ term of $\ket{q_{p_{3}}\bar{q}_{p_{4}}:\text{coh}}$ is given by Eq.(\ref{eq:qqbar_coh_st_V_1}). Inserting this in the third term of Eq.(\ref{eq:pert_expn_reqd}) leads to the following contribution to $T_{fi}$:
\begin{equation} \label{eq:T_3}
\begin{split}
T_{3}\equiv &\mel{q_{p_{3}}\bar{q}_{p_{4}}}{\Omega_{\Delta(V_{1})}V_{1}\frac{1}{p^{-}_{i}-H_{0}}V_{em}\frac{1}{p^{-}_{i}-H_{0}}V_{em}}{e^{-}_{p_{1}}e^{+}_{p_{2}}}\\
=&\ T_{3a}+T_{3b}+T_{3c}+T_{3d},
\end{split}
\end{equation}
where
\begin{equation} \label{eq:T_3a}
\begin{split}
T_{3a}=& -\frac{e^{2}g^{2}}{(2\pi)^{6}}\int\frac{d^{2}{\bf{k}}_{\perp}dk^{+}\Theta_{\Delta}}{\prod_{i}^{}\sqrt{2p^{+}_{i}}}\Bigg[\frac{\bar{u}_{p_{3}}\slashed{\epsilon}^{a}_{k}u_{(p_{3}-k)}\bar{u}_{(p_{3}-k)}\slashed{\epsilon}^{*a}_{k}u_{p_{3}}\bar{u}_{p_{3}}}{2(p^{+}_{1}+p^{+}_{2})2k^{+}2p^{+}_{3}2(p^{+}_{3}-k^{+})}\\
& \frac{\slashed{\epsilon}_{(p_{1}+p_{2})}v_{p_{4}}\bar{v}_{p_{2}}\slashed{\epsilon}^{*}_{(p_{1}+p_{2})}u_{p_{1}}C_{F}\delta_{c_{3}c_{4}}}{(p^{-}_{3}-k^{-}-(p_{3}-k)^{-})(p^{-}_{1}+p^{-}_{2}-p^{-}_{3}-p^{-}_{4})(p^{-}_{1}+p^{-}_{2}-(p_{1}+p_{2})^{-})}\Bigg]
\end{split}
\end{equation}
\begin{equation} \label{eq:T_3b}
\begin{split}
T_{3b}=& -\frac{e^{2}g^{2}}{(2\pi)^{6}}\int\frac{d^{2}{\bf{k}}_{\perp}dk^{+}\Theta_{\Delta}}{\prod_{i}^{}\sqrt{2p^{+}_{i}}}\Bigg[\frac{\bar{u}_{p_{3}}\slashed{\epsilon}_{(p_{1}+p_{2})}v_{p_{4}}\bar{v}_{p_{4}}\slashed{\epsilon}^{*a}_{k}v_{(p_{4}-k)}\bar{v}_{(p_{4}-k)}\slashed{\epsilon}^{a}_{k}}{2(p^{+}_{1}+p^{+}_{2})2k^{+}2p^{+}_{4}2(p^{+}_{4}-k^{+})}\\
& \frac{v_{p_{4}}\bar{v}_{p_{2}}\slashed{\epsilon}^{*}_{(p_{1}+p_{2})}u_{p_{1}}C_{F}\delta_{c_{3}c_{4}}}{(p^{-}_{4}-k^{-}-(p_{4}-k)^{-})(p^{-}_{1}+p^{-}_{2}-p^{-}_{3}-p^{-}_{4})(p^{-}_{1}+p^{-}_{2}-(p_{1}+p_{2})^{-})}\Bigg]
\end{split}
\end{equation}
\begin{equation} \label{eq:T_3c}
\begin{split}
T_{3c}=& \frac{e^{2}g^{2}}{(2\pi)^{6}}\int\frac{d^{2}{\bf{k}}_{\perp}dk^{+}\Theta_{\Delta}}{\prod_{i}^{}\sqrt{2p^{+}_{i}}}\Bigg[\frac{\bar{u}_{p_{3}}\slashed{\epsilon}^{*a}_{k}u_{(p_{3}+k)}\bar{u}_{(p_{3}+k)}\slashed{\epsilon}_{(p_{1}+p_{2})}v_{(p_{4}-k)}\bar{v}_{(p_{4}-k)}}{2(p^{+}_{1}+p^{+}_{2})2k^{+}2(p^{+}_{3}+k^{+})2(p^{+}_{4}-k^{+})}\\
& \frac{\slashed{\epsilon}^{a}_{k}v_{p_{4}}\bar{v}_{p_{2}}\slashed{\epsilon}^{*}_{(p_{1}+p_{2})}u_{p_{1}}C_{F}\delta_{c_{3}c_{4}}}{(p^{-}_{4}-k^{-}-(p_{4}-k)^{-})(p^{-}_{1}+p^{-}_{2}-(p_{3}+k)^{-}-(p_{4}-k)^{-})(p^{-}_{1}+p^{-}_{2}-(p_{1}+p_{2})^{-})}\Bigg]
\end{split}
\end{equation}
\begin{equation} \label{eq:T_3d}
\begin{split}
T_{3d}=& \frac{e^{2}g^{2}}{(2\pi)^{6}}\int\frac{d^{2}{\bf{k}}_{\perp}dk^{+}\Theta_{\Delta}}{\prod_{i}^{}\sqrt{2p^{+}_{i}}}\Bigg[\frac{\bar{u}_{p_{3}}\slashed{\epsilon}^{a}_{k}u_{(p_{3}-k)}\bar{u}_{(p_{3}-k)}\slashed{\epsilon}_{(p_{1}+p_{2})}v_{(p_{4}+k)}\bar{v}_{(p_{4}+k)}}{2(p^{+}_{1}+p^{+}_{2})2k^{+}2(p^{+}_{3}-k^{+})2(p^{+}_{4}+k^{+})}\\
& \frac{\slashed{\epsilon}^{*a}_{k}v_{p_{4}}\bar{v}_{p_{2}}\slashed{\epsilon}^{*}_{(p_{1}+p_{2})}u_{p_{1}}C_{F}\delta_{c_{3}c_{4}}}{(p^{-}_{3}-k^{-}-(p_{3}-k)^{-})(p^{-}_{1}+p^{-}_{2}-(p_{3}-k)^{-}-(p_{4}+k)^{-})(p^{-}_{1}+p^{-}_{2}-(p_{1}+p_{2})^{-})}\Bigg].
\end{split}
\end{equation}
The $\Theta_{\Delta}$ function that appears in the expressions for $T_{3a}$, $T_{3b}$, $T_{3c}$ and $T_{3d}$ given by Eqs.(\ref{eq:T_3a})-(\ref{eq:T_3d}) restricts the region of phase space and hence the integration region to only those values for which the following conditions are satisfied.\\
$T_{3a}$ : $(p^{-}_{3}-k^{-}-(p_{3}-k)^{-})\rightarrow 0$\\
$T_{3b}$ : $(p^{-}_{4}-k^{-}-(p_{4}-k)^{-})\rightarrow 0$\\
$T_{3c}$ : $(p^{-}_{4}-k^{-}-(p_{4}-k)^{-})\rightarrow 0$; $(p^{-}_{1}+p^{-}_{2}-(p_{3}+k)^{-}-(p_{4}-k)^{-}\rightarrow 0$\\
$T_{3d}$ : $(p^{-}_{3}-k^{-}-(p_{3}-k)^{-})\rightarrow 0$; $(p^{-}_{1}+p^{-}_{2}-(p_{4}+k)^{-}-(p_{3}-k)^{-})\rightarrow 0$.\\
The diagrams corresponding to $T_{3}$ are represented schematically by Fig.\ref{fig:regular_coh_diagrams}, in which the dashed lines represent emission/absorption of a soft/collinear gluon. These particles, lying in the restricted phase space through the $\Theta_{\Delta}$ function, are indistinguishable from the final-state particles.

The last term of Eq.(\ref{eq:pert_expn_reqd}) involves the $\mathcal{O}(g^{2})$ contribution to the coherent state $\ket{q_{p_{3}}\bar{q}_{p_{4}}:\text{coh}}$. The expression for this term is
\begin{equation} \label{eq:T_4}
\begin{split}
T_{4}\equiv &\mel{q_{p_{3}}\bar{q}_{p_{4}}}{\Omega_{\Delta(W_{1})}V_{em}\frac{1}{p^{-}_{i}-H_{0}}V_{em}}{e^{-}_{p_{1}}e^{+}_{p_{2}}}\\
=& \frac{e^{2}g^{2}}{(2\pi)^{6}}\int\frac{d^{2}{\bf{k}}_{\perp}dk^{+}\Theta_{\Delta}}{\prod_{i}^{}\sqrt{2p^{+}_{i}}}\Bigg[\frac{\bar{u}_{p_{3}}\gamma^{+}u_{k}\bar{u}_{k}\slashed{\epsilon}_{(p_{1}+p_{2})}v_{(p_{3}+p_{4}-k)}\bar{v}_{(p_{3}+p_{4}-k)}}{2(p^{+}_{1}+p^{+}_{2})2k^{+}2(p^{+}_{3}+p^{+}_{4}-k^{+})(p^{+}_{3}-k^{+})^{2}}\\
& \frac{\gamma^{+}v_{p_{4}}\bar{v}_{p_{2}}\slashed{\epsilon}^{*}_{(p_{1}+p_{2})}u_{p_{1}}C_{F}\delta_{c_{3}c_{4}}}{(p^{-}_{3}+p^{-}_{4}-k^{-}-(p_{3}+p_{4}-k)^{-})(p^{-}_{1}+p^{-}_{2}-(p_{1}+p_{2})^{-})}\Bigg].
\end{split}
\end{equation}
$\Theta_{\Delta}$ restricts the region of integration to only those values where $(p^{-}_{3}+p^{-}_{4}-k^{-}-(p_{3}+p_{4}-k)^{-})\rightarrow 0$.

$T_{fi}$ is thus the sum total of Eqs.(\ref{eq:T_1a})-(\ref{eq:T_2}) and (\ref{eq:T_3a})-(\ref{eq:T_4}). The regions of phase space in which $T_{3a}$ to $T_{3d}$ and $T_{4}$ are nonvanishing are the same which lead to IR divergences in $T_{1a}$ to $T_{1d}$ and $T_{2}$ respectively. The expressions being equal and opposite, the IR divergences cancel and the total amplitude $T_{fi}= \mel{q_{p_{3}}\bar{q}_{p_{4}}:\text{coh}}{T}{e^{-}_{p_{1}}e^{+}_{p_{2}}}$ is IR-finite.

\section{Summary and Outlook} \label{sec:summary}
We show that in LFQCD the total amplitude $T_{fi}$ for the process $e^{+}e^{-}\rightarrow 2\; \text{jets}$ to $\mathcal{O}(g^{2})$ is free of IR divergences. After reviewing the derivation of LFQCD Hamiltonian, we take a brief look at IR divergences appearing in LFQCD. The coherent state formalism is then described and the outgoing coherent state required for calculating the $e^{+}e^{-}\rightarrow 2\; \text{jets}$ amplitude is constructed. Noting that vanishing energy denominators in the amplitude in Fock basis lead to soft/collinear divergences, we show that these divergences get eliminated when higher order coherent state contributions are added to the amplitude. The cancellation of IR divergences takes place through $\Theta_{\Delta}$ function, which restricts the integration to only that region of phase space which leads to IR divergences. Hence, to $\mathcal{O}(g^{2})$, we show that the total amplitude $T_{fi}=\mel{q_{p_{3}}\bar{q}_{p_{4}}:\text{coh}}{T}{e^{-}_{p_{1}}e^{+}_{p_{2}}}$ is free of soft and collinear divergences and thus IR-finite.

Parton showering algorithms that are currently used for studying high-energy scattering phenomena employ the conventional technique of cancellation of IR divergences at the cross-section level. Recently, construction of all-order amplitude-level parton-shower algorithms is being attempted \cite{martinez}. The coherent state approach is expected to play an important role in such algorithms by providing a way to deal with IR divergences at the amplitude level itself. Keeping this aspect in view, the preliminary study carried out in this work needs to be taken further. Our followup work will deal with cancellation of IR divergences, in the coherent state formalism, in multijet events which are now being regularly observed in the currently operating colliders. Higher-order calculations will also be carried out in the near future with the aim of developing an all-order proof of cancellation of IR divergences in pQCD.

\section*{Acknowledgments}
D.B. would like to acknowledge the financial support provided by University Grants Commission (UGC), India.
A.M. would like to thank ICTP for their kind hospitality and support under the Senior Associate program.

\appendix 
\section{Amplitude $T_{3a}$} \label{app:Amplitude_calculation}
The details of the calculation for the expression of amplitude $T_{3a}$ are presented below. The amplitude $T_{3a}$ corresponds to the diagram of Fig.\ref{fig:regular_coh_diagrams}(a). $T_{3}$ is given by Eq.(\ref{eq:T_3}).
Inserting complete sets of states between operators and writing only those states of the complete set which give a nonvanishing contribution, the amplitude becomes 
\begin{equation} \label{eq:states_inserted}
\begin{split}
T_{3}=& \int d^{3}kd^{3}k'\prod_{i=1}^{2}d^{3}q_{i}\prod_{i=1}^{2}d^{3}q'_{i}\mel{q_{p_{3}}\bar{q}_{p_{4}}}{\Omega_{\Delta(V_{1})}V_{1}}{q_{q_{1}}\bar{q}_{q_{2}}}\mel{q_{q_{1}}\bar{q}_{q_{2}}}{\frac{1}{p^{-}_{1}+p^{-}_{2}-H_{0}}}{q_{q'_{1}}\bar{q}_{q'_{2}}}\\
& \mel{q_{q'_{1}}\bar{q}_{q'_{2}}}{V_{em}}{\gamma_{k}}\mel{\gamma_{k}}{\frac{1}{p^{-}_{1}+p^{-}_{2}-H_{0}}}{\gamma_{k'}}\mel{\gamma_{k'}}{V_{em}}{e^{-}_{p_{1}}e^{+}_{p_{2}}},
\end{split}
\end{equation}
where $d^{3}p\equiv d^{2}{\bf{p}}_{\perp}dp^{+}$.\\
Using the identities 
\begin{equation}
\begin{split}
& \braket{\gamma_{k}}{\gamma_{k'}}=\delta(k-k')\\
& \braket{q_{q_{1}}\bar{q}_{q_{2}}}{q_{q'_{1}}\bar{q}_{q'_{2}}}=\delta(q_{1}-q'_{1})\delta(q_{2}-q'_{2})
\end{split}
\end{equation}
to eliminate the momenta $k'$, $q'_{1}$ and $q'_{2}$ (and other corresponding quantum numbers) from Eq.(\ref{eq:states_inserted}) reduces the amplitude to
\begin{equation}
T_{3}= \int d^{3}kd^{3}q_{1}d^{3}q_{2}\frac{\mel{q_{p_{3}}\bar{q}_{p_{4}}}{\Omega_{\Delta(V_{1})}V_{1}}{q_{q_{1}}\bar{q}_{q_{2}}}\mel{q_{q_{1}}\bar{q}_{q_{2}}}{V_{em}}{\gamma_{k}}\mel{\gamma_{k}}{V_{em}}{e^{-}_{p_{1}}e^{+}_{p_{2}}}}{({p^{-}_{1}+p^{-}_{2}-q^{-}_{1}-q^{-}_{2}})({p^{-}_{1}+p^{-}_{2}-k^{-}})}.
\end{equation}
Consider the inner product $\mel{\gamma_{k}}{V_{em}}{e^{-}_{p_{1}}e^{+}_{p_{2}}}$. Substituting for $V_{em}=e\int d^{2}{\bf{x}}_{\perp}dx^{-} \bar{\Psi}\gamma^{\mu}\Psi A_{\mu}$, we get
\begin{equation} \label{eq:inner_prod_oprs}
\begin{split}
\mel{\gamma_{k}}{V_{em}}{e^{-}_{p_{1}}e^{+}_{p_{2}}}=& e\int d^{2}{\bf{x}}_{\perp}dx^{-}\int \prod_{i}^{}[dk_{i}]\bra{0}a_{k}(b^{\dagger}_{k_{1}}\bar{u}_{k_{1}}e^{ik_{1}{\cdot}x}+d_{k_{1}}\bar{v}_{k_{1}}e^{-ik_{1}{\cdot}x})\\
& \gamma^{\mu}(b_{k_{2}}u_{k_{2}}e^{-ik_{2}{\cdot}x}+d^{\dagger}_{k_{2}}v_{k_{2}}e^{ik_{2}{\cdot}x})(a_{k_{3}}\epsilon^{a}_{\mu k_{3}}e^{-ik_{3}{\cdot}x}\\
&+ a^{\dagger}_{k_{3}}\epsilon^{*a}_{\mu k_{3}}e^{ik_{3}{\cdot}x})b^{\dagger}_{p_{1}}d^{\dagger}_{p_{2}}\ket{0}.
\end{split}
\end{equation}
The only nonvanishing string of creation/annihilation operators gives $\mel{0}{a_{k}d_{k_{1}}b_{k_{2}}a^{\dagger}_{k_{3}}b^{\dagger}_{p_{1}}d^{\dagger}_{p_{2}}}{0}=\delta(k_{3}-k)\delta(k_{2}-p_{1})\delta(k_{1}-p_{2})$. Integrating over momenta $k_{i}$ in Eq.(\ref{eq:inner_prod_oprs}) using these delta functions, we obtain
\begin{equation} \label{eq:inner_prod_exprn}
\mel{\gamma_{k}}{V_{em}}{e^{-}_{p_{1}}e^{+}_{p_{2}}}=\frac{e\bar{v}_{p_{2}}\gamma^{\mu}u_{p_{1}}\epsilon^{*}_{\mu k}\delta(p_{1}+p_{2}-k)}{(2\pi)^{3/2}\sqrt{2p^{+}_{1}}\sqrt{2p^{+}_{2}}\sqrt{2k^{+}}}.
\end{equation}
Similarly,
\begin{equation}
\mel{q_{q_{1}}\bar{q}_{q_{2}}}{V_{em}}{\gamma_{k}}=\frac{e\bar{u}_{q_{1}}\gamma^{\mu}v_{q_{2}}\epsilon_{\mu k}\delta(q_{1}+q_{2}-k)}{(2\pi)^{3/2}\sqrt{2q^{+}_{1}}\sqrt{2q^{+}_{2}}\sqrt{2k^{+}}}.
\end{equation}\\
The (Hermitian conjugate of) outgoing coherent state $\Omega^{\dagger}_{\Delta(V_{1})}\ket{q_{p_{3}}\bar{q}_{p_{4}}}$ given by Eq.(\ref{eq:qqbar_coh_st_V_1})
is substituted in $\mel{q_{p_{3}}\bar{q}_{p_{4}}}{\Omega_{\Delta(V_{1})}{V_{1}}}{q_{q_{1}}\bar{q}_{q_{2}}}$. This leads to four nonvanishing terms, out of which the one containing the string $d_{p_{4}}b_{(p_{3}-k)}a_{k}b^{\dagger}_{n_{1}}d^{\dagger}_{n_{2}}a^{\dagger}_{n_{3}}b^{\dagger}_{q_{1}}d^{\dagger}_{q_{2}}$, where $n_{i}$ are momenta labels for fields appearing in the interaction $V_{1}$, eventually gives the amplitude $T_{3a}$. The other three terms give the three remaining amplitudes $T_{3b-3d}$. $T_{3a}$ is now obtained by performing momentum integrations using delta functions, and is given by Eq.(\ref{eq:T_3a}).

\bibliography{references}

\end{document}